\documentclass[aps,prl,letterpaper,amsmath,amssymb,superscriptaddress,nofootinbib,twocolumn]{revtex4-2}

\usepackage{graphicx}
\usepackage[usenames]{color}
\usepackage[colorlinks=true,linkcolor=magenta,citecolor=magenta,anchorcolor=green,urlcolor=magenta]{hyperref}
\usepackage{mathrsfs}

\usepackage[percent]{overpic}
\usepackage{tabularx}
\usepackage{multirow}
\usepackage{orcidlink}

\usepackage{CJK}
\newcommand\w           {\omega}
\newcommand{\wec}[1]   {|#1\rangle}
\newcommand{\cew}[1]   {\langle#1|}

\newcommand{\schro} {\text{Schr\"{o}dinger}}

\def\be{\begin{equation}}
\def\ee{\end{equation}}
\def\ba{\begin{aligned}}
\def\ea{\end{aligned}}
\def\bea{\begin{eqnarray}}
\def\eea{\end{eqnarray}}

\begin{document}
\begin{CJK*}{UTF8}{}
\title{Berry connection and quantum geometry in time-dependent systems with instantaneous quantum integrable field theory}
\author{Xiao Wang \CJKfamily{gbsn}(王骁)~\orcidlink{0000-0003-2898-3355}}
\affiliation{Tsung-Dao Lee Institute,
Shanghai Jiao Tong University, Shanghai, 201210, China}

\author{Xiaodong He \CJKfamily{gbsn}(何晓东)~\orcidlink{0009-0008-4724-9483}}
\affiliation{Tsung-Dao Lee Institute,
Shanghai Jiao Tong University, Shanghai, 201210, China}

\author{Jianda Wu \CJKfamily{gbsn}(吴建达)~\orcidlink{0000-0002-3571-3348}}
\altaffiliation{wujd@sjtu.edu.cn}
\affiliation{Tsung-Dao Lee Institute,
Shanghai Jiao Tong University, Shanghai, 201210, China}
\affiliation{School of Physics \& Astronomy, Shanghai Jiao Tong University, Shanghai, 200240, China}
\affiliation{Shanghai Branch, Hefei National Laboratory, Shanghai 201315, China}

\begin{abstract}
We study many-body quantum geometric effects in time-dependent system with
emergent quantum integrable field theory instantaneously.
We establish a theorem stating that the Berry connection matrix thus 
all associated geometric and topological quantities 
of the system can be precisely characterized by excitations up to two particles 
from the initial emergent quantum integrable field theory. 
To illustrate the many-body geometric influence, we analyze an Ising chain subjected 
to both a small longitudinal field and a slowly rotating transverse field, whose low-energy physics in the scaling limit 
is instantaneously governed by the quantum $E_8$ integrable field theory. 
Focusing on the quantum geometric potential (QGP), we show the QGP continuously suppresses 
the instantaneous energy gaps with decreasing longitudinal field, 
thereby enhancing many-body Landau-Zener tunneling as evidenced by the Loschmidt echo 
and its associated spectral entropy. 
The critical threshold for the longitudinal field strength is determined,
where the spectral entropy linearly increases with system size and exhibits
hyperscaling behavior when approaching to the threshold.
As the longitudinal field passes the threshold and decreases toward zero, 
the QGP continuously leads to vanishing instantaneous 
energy gaps involving more low-energy excitations, 
resulting in increasing spectral entropy indicative of many-body Landau-Zener tunneling.
Our results unveil telltale quantum geometric signatures in time-dependent many-body systems, 
elucidating the intricate interplay between quantum geometry, integrability, and dynamics.
\end{abstract}

\maketitle
\end{CJK*}

\textit{Introduction}.---
Quantum geometric effects have attracted sustained 
research interests over the past several decades since the introduction of the Berry phase~\cite{berryphase}. 
Significant progress has been made in fields such as topological band insulators~\cite{guangyu_prx,guangyu_np,gy_prb2023}, 
non-equilibrium adiabatic evolution~\cite{da_tunneling,RMP_adiabatic}, and the quantum Hall effect~\cite{PhysRevLett.49.405}. 
For instance, the higher-order optical responses can be generally characterized by 
the Riemannian geometry of quantum states~\cite{guangyu_prx,guangyu_np,gy_prb2023}.
Despite the clear structure of quantum geometric quantities constructed from wavefunctions, 
studying quantum geometric effects in quantum many-body systems remains highly challenging, 
and related research is still scarce~\cite{korepin_1991}. This scarcity primarily stems 
from the complexity of many-body wavefunctions which involve an exponentially growing 
Hilbert space in such systems. For instance, calculating Berry connections in many-body systems poses significant challenges, both numerically and analytically.

Due to exact solvability, quantum integrable systems may provide critical insights into quantum geometry in many-body systems:
Their eigenstates can be completely
labeled by the corresponding conserved quantum numbers~\cite{LL_I_1963,LL_II_1963,GAUDIN1967,CNYang_1967,Jimbo1995,Guan_Fermi_2013,He_quantum_2017,Franchini2017,yang_magnetic_2023,yjh_2024, Korepin_1993,BM_2004}, 
therefore provide a concrete base to 
analytically study quantum geometry in quantum integrable systems~\cite{smirnov_1992,delfino_1995,kh2020,PhysRevLett.52.2111,da_QGP2008, da_QGP2018}.
Over the past decade, significant research has focused on quantum integrability in non-equilibrium systems, including integrable quantum quenches and integrable brickwork circuits~\cite{delfino_2014,delfino_2024,PhysRevLett.110.257203,CalabreseEsslerFagotti,marton2016,Delfino_2017,kh_2018,Etienne2020,PhysRevLett.121.030606,PhysRevLett.130.260401,yunfeng_2024,xiao_2024time,DiSalvo}. These studies have provided new insights into non-equilibrium physics from the perspective of integrability, with potential connections to emerging fields such as quantum computing and quantum information. However, the understanding of quantum geometric effects in quantum many-body (integrable) systems remains underdeveloped~\cite{Paul_1992}.

To study quantum geometry and its effects in
quantum many-body systems, 
we focus on time-dependent systems with emergent quantum integrable field theory instantaneously.
The $n$-th instantaneous eigenfunction is $\vert \phi_n(t)\rangle$ with instantaneous eigenenergy $e_n(t)$ at the moment $t$.
After imposing a proper constraint on the adiabatic transformation $\mathscr{U}_a(t): \vert{\phi}_n(0)\rangle \to \vert{\phi}_n(t)\rangle $, 
we prove that the Berry connection matrix (BCM) entries 
$\gamma_{nm} = i \langle\phi_n(t)\vert\dot{\phi}_m(t)\rangle$ only associate excitations up to two particle processes.

For illustration we further consider the time evolution of 
a quantum Ising chain under both a slowly rotating transverse field and a small static longitudinal 
field~\cite{Robinson,xiao_2024time}, where the initial state is set as the ground state at the beginning.
We introduce a local U(1) gauge-invariant
geometric term known as the quantum geometric potential (QGP) $Q_{mn} = \gamma_{mm} - \gamma_{nn} + \frac{d}{dt} \text{arg}~\gamma_{nm}$ ~\cite{da_QGP2008,da_QGP2018} generated from any two instantaneous eigenfunctions. This term
provides a crucial correction to
the instantaneous energy gap $e_{nm} = e_n (t) - e_m (t)$ 
and thus gives rise to
a more efficient
effective energy gap (EEG)
$\Delta_{nm} =e_{nm} + Q_{mn}$. Notably,
even when the instantaneous energy gap $e_{nm}$ is large, 
perfect Landau-Zener tunneling can still occur if the $\Delta_{nm}$ vanishes
when the QGP cancels the $e_{nm}$. 
This has been 
validated in the NMR experiment~\cite{DU_2008} and further
demonstrated in time-dependent two-band models~\cite{da_tunneling}.

In the scaling limit, the QGP of this system can be analytically determined through the quantum $E_8$ integrable field theory (IFT). 
The EEG can then be reduced to a more compact formalism as a function of the rotating frequency ($\omega_0$) and the longitudinal field intensity ($h_z$). 
We first identify our calculation of the QGP with the truncated conformal space approach (TCSA).

Then to show the influence of the QGP and EEG in many-body Landau-Zener tunneling (MBLZT) processes, 
we employ the truncated lattice free fermion approach (TLFFA)~\cite{SM,Iorgov2011,zam_1990,Tu2410}. 
Within this framework, we analyze the spectral entropy of the Loschmidt echo (LE), 
which quantifies the overlap between the time-evolved wavefunction and the instantaneous ground state for different system sizes $L$. 
We find a critical $\kappa \equiv h_z/(\omega_0/2) $, 
 $\kappa_c = 1$ which serves as a critical threshold for the MBLZT. 
When $\kappa > 1$, zeros of EEG are nearly absent, manifested as normal spectral entropy contributed from a limited number of finite-frequency modes.
As $\kappa$ goes to 1, a hyperscaling behavior of spectral entropy appears according to the scaling parameter $(\kappa-1)^{8/15}L$. 
At $\kappa=1$, the spectral entropy scales linearly with the system size $L$,
implying MBLZT avalanche in thermodynamic limit.
When $\kappa \leq 1$ zeros of EEG  substantially increase with decreasing $\kappa$, 
manifested as persistent growth of the LE spectral entropy,
signaling significantly enhanced MBLZT.
Interestingly, the LE spectral entropy also exhibits a cascade of additional
peaks when $\kappa \leq 1$, where we expect strengthened MBLZT.

\begin{figure}[tp]
    \includegraphics[width=6cm]{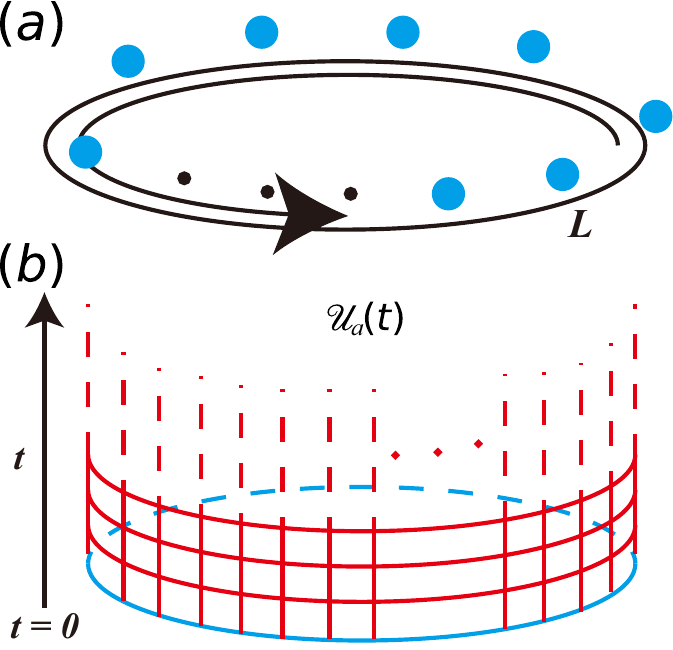}
    \caption{An illustration of the setup for the theorem.
(a) The initial wavefunction of an eigenstate in an IFT contains $N$ quasiparticles with periodic boundary conditions, represented as a cylinder with circumference $L$. The blue circles represent the quasiparticles.
(b) The time evolution of the initial wavefunction under the unitary transformation $\mathscr{U}_a(t)$, with the effective adiabatic Hamiltonian given by Eq.~(\ref{Eq:Hadia}). The blue line corresponds to the initial configuration in (a), and the red space-time network illustrates $\mathscr{U}_a(t)$.}
    \label{fig:illustration}
\end{figure}

\begin{figure*}[tp]
    \includegraphics[width=17cm]{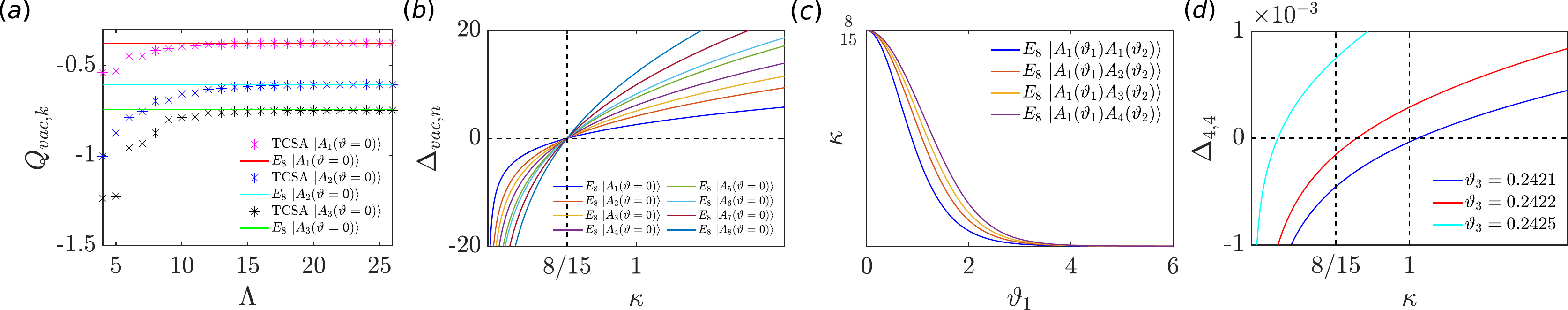}
    \caption{(a) Comparison of the QGP between the ground state (vacuum of the $\mathcal{H}_{E_8}$) and the three lightest 
    single-particle states of the $\mathcal{H}_{E_8}$, calculated with both the $\mathcal{H}_{E_8}$ form factor theory and the 
    boosted TCSA algorithm ~\cite{TCSA}. Here, $\Lambda$ is the truncation parameter satisfying $E^2-P^2\leq \Lambda^2$. 
    (b) Analytical results of the EEG between the ground state and the 8 single $E_8$ particle state with zero momentum, 
    with $0<\kappa \le 2$. The zeros of EEG appear at $\kappa=8/15$ 
    for all the single $E_8$ particles. (c) The zeros of EEG as a function of $A_1$'s rapidity $\vartheta_1$, for the two-particle states $A_1A_1$ to $A_1A_4$, with a zero momentum constraint $m_{i}\sinh\vartheta_1+m_{j}\sinh\vartheta_2=0$. (d) Zeros of EEG between two four $E_8$ particle states. While both the two groups of particles are of particle type $A_{1\sim4} $, the rapidities for one of the states takes $(\vartheta_1,\vartheta_2,\vartheta_3,\vartheta_4)=(0.2,0.2,0.1,0.1)$, and for the other one takes $\vartheta_1=\vartheta_2=\vartheta_4=0.1$ and $\vartheta_3 = 0.2421,0.2422,0.2425$, corresponding to the zeros being from large to small, respectively.}
    \label{fig:QGPandEEG_E8tcsa}
\end{figure*}

\textit{The BCM in instantaneously quantum-integrable time-dependent systems}.---
It is generally difficult to calculate the BCM in interacting quantum many-body systems. 
We consider time-dependent quantum system with emergent 1+1D quantum field theory instantaneously, 
where the initial state is an eigenstate in an IFT containing $N$ stable quasiparticles [Fig.~\ref{fig:illustration}(a)], 
and the instantaneous eigenstate is connected to the initial state 
with an adiabatic transformation, as $\wec{\phi_{N}(t)}=\mathscr{U}_a(t)\wec{\phi_{N}(0)}_{\text{init},L}$ [Fig.~\ref{fig:illustration}(b)].
Without loss of generality the system is placed on a cylinder with 
circumference $L$. We require the effective adiabatic Hamiltonian~\cite{yongdezhang} 
$\mathcal{H}_{adia}=i\mathscr{U}_a^{-1}(t)\dot{\mathscr{U}_a}(t)$ to satisfy 
\begin{equation}
\mathcal{H}_{adia}=\sum_{i=1}^{N_{\mathcal{O}}}\int_{0}^{L}f_{i}(t)\mathcal{O}_{i}(x)dx
\label{Eq:Hadia}
\end{equation}
in the continuum limit. Here $\mathcal{O}_{i}(x)$ are the local field densities. 
The initial state is given by
$\wec{\phi_{N}(0)}_{\text{init},L}=\wec{A_{1}(\vartheta_1;I_1)...A_{N}(\vartheta_N;I_N)}_{L}$.
Here, $A_{j},j=1,...,N$ label different types of particles with corresponding mass $m_j$, discrete 
quantum number $I_{j}$, and rapidity $\vartheta_{j}$. In addition, the energy and 
momentum of this state are given by $E=\sum_{j}m_{j}\cosh\vartheta_j$ and $P=\sum_{j}m_{j}\sinh\vartheta_j$. 
For any single particle $A_{j}(\vartheta_j;I_j)$, the discrete momentum and rapidity are connected with each other by the Bethe-Yang quantization condition derived from the spatial periodic boundary condition~\cite{POZSGAY2008167,POZSGAY2008209,H_ds_gi_2019,xiao_2024time,SM}.
The BCM can then be explicitly given with the help of quantum integrability. 
Considering two such states containing $N$ and $M$ quasiparticles respectively, for the off-diagonal matrix element where $M\neq N$, or $M=N$ but the two sets of rapidities are not exactly same, the BCM is given by
\begin{equation}
    \frac{\gamma_{NM}}{L}\approx \sum_{i}f_i(t)\frac{\mathcal{F}_{NM}^{\mathcal{O}_i}(\{\vartheta'+i\pi\}_{N},\{\vartheta\}_{M})}{\sqrt{\rho_{L,\{\mathscr{A'}\}_{N}}}\sqrt{\rho_{L,\{\mathscr{A}\}_{M}}}},
\end{equation}
with L\"{u}scher's remainder $o(e^{-\mu L})$~\cite{mass,POZSGAY2008167,POZSGAY2008209,H_ds_gi_2019}. 
Here, $\{\vartheta'+i\pi\}_{N}=\{\vartheta'_{1}+i\pi,...,\vartheta'_{N}+i\pi$\} and $\{\vartheta\}_{M}=\{\vartheta_{1},...,\vartheta_{M}\}$ 
are the sets of rapidities of the two particle sets $\{\mathscr{A'}\}_{N}=\{A'_{1},...,A'_{N}\}$ and $\{\mathscr{A}\}_{M}=\{A_{1},...,A_{M}\}$.
The function $\mathcal{F}_{NM}^{\mathcal{O}_i}(\{\vartheta'+i\pi\}_{N},\{\vartheta\}_{M})$ 
in terms of rapidities is the matrix element of operator $\mathcal{O}_{i}$ with respect to the two multi-particle states, 
known as the form factor of IFT~\cite{SM,smirnov_1992,BM_2004,delfino_1995,kh2020}.
The two functions of $\rho_{L,\{\mathscr{A'}\}_{N}}$ and $\rho_{L,\{\mathscr{A}\}_{M}}$ are Jacobian
from the particle density of states of the two particle sets 
in the finite size system,
where the detailed expressions can be found in~\cite{SM}. 
For the diagonal matrix element where both $M=N$, and the two sets of rapidities are exactly the same, the result is given by
\begin{equation}
\begin{aligned}
    \frac{\gamma_{NN}}{L}&\approx\sum_{i}f_{i}(t)\bigg{[}{}_{L}\langle\mathcal{O}_{i}\rangle_{L}\\
    &+\sum_{\forall \mathscr{S}\subseteq \{\mathscr{A}\}_{N}}\frac{\mathcal{F}_{\mathscr{S}\mathscr{S}}^{\mathcal{O}_{i}}(\{\vartheta+i\pi\}_{\mathscr{S}},\{\vartheta\}_{\mathscr{S}})\rho_{L,\{\mathscr{A}\}_{N}\setminus \mathscr{S}}}{\rho_{L,\{\mathscr{A}\}_{N}}}\bigg{]},
\end{aligned}
\end{equation}
with the L\"{u}scher's remainder as well. 
Here ${}_{L}\langle\mathcal{O}_{i}\rangle_{L}$ corresponds to the vacuum expectation value of $\mathcal{O}_{i}$,
$\mathscr{S}$ is non-empty subset of $\{\mathscr{A}\}_{N}$. 
$\{\mathscr{A}\}_{N}\setminus \mathscr{S}$ denotes particle set excluding $\mathscr{S}$ from $\{\mathscr{A}\}_{N}$.
The full information of the BCM implies that 
for such quantum many-body systems, we can determine all the geometric quantities 
associated with the BCM. Notably, the Jacobians in the denominator provide 
suppression behavior of order $L^{\mathcal{Z}}$ with total number of particle $\mathcal{Z} = M + N$. 
The algebraic decaying behaviour of the Jacobians in the limit of $L\rightarrow\infty$~\cite{SM} 
consequently leads to the asymptotic behaviour of the BCM elements. Thus we prove the following theorem.

\textbf{Theorem:} \textit{Consider time-dependent many-body quantum systems 
with emergent quantum integrable field theory instantaneously,
if the effective adiabatic Hamiltonian follows Eq.~(\ref{Eq:Hadia}), then
the corresponding BCM entries with total quasi-particle numbers of $\mathcal{Z}$ follow 
the asymptotic algebraic behaviour $\sim L^{1-\mathcal{Z}/2}$ as the system size $L\rightarrow\infty$.}

This theorem suggests that in the thermodynamic limit ($L\rightarrow\infty$) the corresponding BCM entries can be 
analytically determined as excitations up to two particles.
This can also be understood from the quasi-particle nature of IFTs: 
While integrating the field operator over the entire system introduces a volume dependence, 
the Jacobian factor arising from the density of particle states ultimately 
suppresses the asymptotic behaviour of the BCM elements for multi-particle states.
In the following, the theorem is applied to a time-dependent quantum Ising model, where the QGP constructed from the BCM entries is shown to play a crucial role in the MBLZT in the system.

\textit{A time-dependent quantum Ising chain with instantaneous quantum $E_8$ integrability}.---
We start from a ferromagnetic Ising chain subjected to both a periodically rotating transverse field and a static longitudinal field ~\cite{Robinson,xiao_2024time},
\begin{equation}
    \mathcal{H}(t)=-\sum_{i=1}^{L}\left(\sigma^{z}_{i}\sigma^{z}_{i+1}+\cos\w_0 t\sigma^{x}_{i}-\sin\w_0 t\sigma^{y}_{i}+h_{z}\sigma^{z}_{i}\right),
    \label{Eq:latH_driven}
\end{equation}
with $\hbar=c=1$, Pauli matrices $\sigma_{i}^{\alpha}(\alpha=x,~y,~z)$
at site $i$, the length of the chain $L$, 
and the neighboring spin coupling $J=1$. Additionally, we require $\w_0$, $h_{z}\ll 1$. 
It is not hard to observe that
the Hamiltonian instantaneously appears as a quantum critical transverse-field Ising chain perturbed by a small longitudinal field, whose low-energy physics in the scaling limit is described by the quantum $E_8$ IFT,
\begin{equation}
\mathcal{H}_{E_8}=\mathcal{H}_{c=1/2}-h\int\sigma(x)dx,
\label{Eq:HE8}
\end{equation}
where $\mathcal{H}_{c=1/2}$ is the central charge $1/2$ conformal field theory, $\sigma(x)$ is known as the spin density operator, and the coupling constant $h$ gives an exact relation with the mass of the lightest particle as $m_1=C_{m}h^{8/15}$, with $C_{m}=4.4049$..., 
and masses of all quantum $E_8$ particles in units of $m_1$ satisfy the eigenvalues of the Cartan matrix for $E_8$ Lie algebra~\cite{fateev,delfino_1995,xiao_2024time,xiao_2021,xiao_2023}.
In the following we set the initial state $\wec{\psi(0)}$ as the ground state of Eq.~(\ref{Eq:latH_driven}) at $t=0$, which in the scaling limit is the vacuum state $\wec{0}_{h}$ of the $\mathcal{H}_{E_8}$. 

The general time-evolved wavefunction of the Hamiltonian Eq.~(\ref{Eq:latH_driven}) follows~\cite{Robinson,xiao_2024time} 
$\wec{\psi(t)}=\mathcal{U}(t)^{-1}e^{-i\mathcal{H}_{\text{eff}}t}\wec{\psi(0)}$,
where the unitary transformation is given by $\mathcal{U}(t)=\exp\left\lbrace -\frac{i}{2}\w_{0} t\sum_{i}\sigma_{i}^{z}\right\rbrace$, satisfying $\mathcal{U}(t)\mathcal{H}(t)\mathcal{U}(t)^{-1}=\mathcal{H}(t=0)$, and the effective Hamiltonian 
$\mathcal{H}_{\text{eff}} = \mathcal{U}\mathcal{H}(t)\mathcal{U}^{-1} + i \dot{\mathcal{U}}\mathcal{U}^{-1}$
follows,
\begin{equation}
\mathcal{H}_{\text{eff}} =-\sum_{i}\left[\sigma_{i}^{z}\sigma_{i+1}^{z}+\sigma_{i}^{x}+(h_{z}-\frac{\w_{0}}{2})\sigma_{i}^{z}\right],
\label{Eq:Heff}
\end{equation}
where for $h_z \neq {\w_{0}}/{2}$, the low-energy physics is described by another $\mathcal{H}_{E_8}$ with modified longitudinal coupling.
Considering the instantaneous \schro~equation $\mathcal{H}(t)\wec{\psi_n(t)}_{ins}=E_{n}(t)\wec{\psi_n(t)}_{ins}$,
where $\wec{\psi_n(t)}_{ins}$ is the instantaneous eigenstate associated
with the $n^{th}$ instantaneous eigenenergy $E_{n}(t)$, it is straightforward to prove~\cite{SM} $E_{n}(t)=E_{n}(0)$ 
and $ \wec{\psi_n(t)}_{ins}=e^{i\varphi_n(t)}\mathcal{U}(t)^{-1}\wec{\psi_n(0)}$, 
which implies that the instantaneous energy levels remain invariant for the Hamiltonian Eq.~(\ref{Eq:latH_driven}), 
and the instantaneous eigenstates are given by a global rotation according to $\mathcal{U}(t)$ up to 
an additional time-dependent overall U(1) gauge $\varphi_{n}(t)$.

Now let's focus on the QGP 
 $Q_{mn}$, taking the scaling limit of the model Eq.~(\ref{Eq:latH_driven}), the QGP follows
\begin{equation}    Q_{nm}=\sum_{j=1}^{M}\frac{\mathcal{F}^{\mathcal{O}}_{jj}(i\pi,0)}{m_{j}\cosh\vartheta_{j}}-\sum_{i=1}^{N}\frac{\mathcal{F}^{\mathcal{O}}_{ii}(i\pi,0)}{m_{i}\cosh\vartheta_{i}}.
    \label{Eq:QGPift}
\end{equation}
As a concrete example, we consider the QGP between the vacuum state and the single $E_8$ particle state with zero momentum, 
which is given by $\mathcal{A}_{00}-\mathcal{A}_{kk}=\langle\sigma\rangle \mathcal{F}^{\sigma}_{kk}(i\pi,0)/m_k$ where $\langle\sigma\rangle=C_{\sigma}h^{1/15}$ with $C_{\sigma}=1.27758\cdots$~\cite{delfino_1995,xiao_2024time}. 
The results agree with those from the boosted TCSA algorithm~\cite{TCSA}, as shown in Fig.~\ref{fig:QGPandEEG_E8tcsa} (a) for $k=1,2,3$.

\textit{The EEG in time-dependent IFT}.---
In the following we consider 
the EEG between two instantaneous eigenstates and try to find the zeros in 
$\Delta_{mn}(t)=0$ for any  $\wec{\psi_{m}}_{ins}$ and $\wec{\psi_{n}}_{ins}$. 
At the zeros, the quantum adiabatic condition 
is significantly violated~\cite{da_QGP2008,DU_2008,da_QGP2018,da_tunneling}, 
and the MBLZT process shall occur. 
Moreover, when $h_z$ decreases the EEG will cross zero and become negative, 
which suggests an effective swap of the high- and low- energy states instantaneously. 
Under this scenario, the MBLZT will be strongly enhanced because more and more higher-energy states effectively 
become low-energy due to the QGP.
In the following we analytically solve the EEG for the "low-energy" sector of the Hamiltonian Eq.~(\ref{Eq:latH_driven}), 
whose results are associated with the $\mathcal{H}_{E_8}$ [Eq.~(\ref{Eq:HE8})].

In the scaling limit, for the low-energy sector the EEG during the time evolution can be solved analytically. 
Considering any two instantaneous eigenstates $\wec{\psi_{n}(t)}_{ins}$ and $\wec{\psi_{m}(t)}_{ins}$ in the model Eq.~(\ref{Eq:latH_driven}), the instantaneous energy gap reads
$E_{mn} = \sum_{\beta=1}^{M}m_{\beta}\cosh\theta_{\beta}-\sum_{\alpha=1}^{N}m_{\alpha}\cosh\theta'_{\alpha}$,
where we have assumed that there are totally $M$ and $N$ quantum $E_8$ particles in the two states, 
respectively. With the mass ratios for the $E_8$ particles $\xi_i$ in units of $m_{1}$, 
the masses of the quantum $E_8$ particles follows $m_{\beta(\alpha)}=\xi_{\beta(\alpha)}m_{1}=\xi_{\beta(\alpha)}C_{m}h^{8/15}_{z}$, 
and $\theta^{(')}_{\beta(\alpha)}$ is the corresponding rapidity. 
The BCM entry is given by $\gamma_{km}(t)=-\dot{\varphi}_{m}(t)\delta_{km}-\w_0/2\int dx\langle\psi_{k}\vert\sigma(x)\vert\psi_{m}\rangle$. 
Applying the quantum $E_8$ form factor theory, the QGP can be organized in a more compact form as~\cite{SM}
\begin{equation}
Q_{nm}(t)=-\frac{8}{15\kappa}\left(\sum_{\beta=1}^{M}\frac{m_{\beta}}{\cosh\theta_{\beta}}-\sum_{\alpha=1}^{N}\frac{m_{\alpha}}{\cosh\theta'_{\alpha}}\right),
    \label{Eq:QGPlattice}
\end{equation}
with $\kappa\equiv h_z/(\w_0/2)$ as introduced in the introduction. Thus the EEG finally reads
\begin{equation}
\begin{aligned}
\Delta_{mn}(t)=&\sum_{\beta=1}^{M}m_{\beta}\cosh\theta_{\beta}-\sum_{\alpha=1}^{N}m_{\alpha}\cosh\theta'_{\alpha}\\
&-\frac{8}{15\kappa}\left(\sum_{\beta=1}^{M}\frac{m_{\beta}}{\cosh\theta_{\beta}}-\sum_{\alpha=1}^{N}\frac{m_{\alpha}}{\cosh\theta'_{\alpha}}\right).
\label{Eq:EEG}
\end{aligned}
\end{equation}
Interestingly, if taking one of the states being the ground state (vacuum) of the initial states, and the other one being the single particle state with zero momentum, the zeros of EEG are given by
$\kappa = 8/15$,
which does not depend on the type of the particles [{\it cf.} Fig.~\ref{fig:QGPandEEG_E8tcsa} (b)]. 
It is also the upper bound for the zeros of EEG between the ground state and the higher excited states~\cite{SM}. 
In the quantum $E_8$ case, the higher excited states are readily given by single particle 
states with large rapidity (momentum) or multi-particle states. We show that for two-particle excitations 
with zero total momentum constraint, as one of the particles' rapidity goes to infinity, 
the corresponding zeros of the EEG will approach 0. It is also easy to verify from Eq.~(\ref{Eq:EEG}) that the zeros 
for two same particle combinations with zero total momentum will collapse to the same curve as $A_1A_1$.
The results are presented in Fig.~\ref{fig:QGPandEEG_E8tcsa} (c). However, if considering the zeros of EEG for any two instantaneous eigenstates, the region for zeros of EEG is beyond $\kappa=8/15$. For instance, Fig.~\ref{fig:QGPandEEG_E8tcsa} (d) shows contribution of four-particle excitations to the EEG whose zeros appear to approach  $\kappa=1$.

\textit{The spectral entropy of the Loschmidt echo}.---
In order to investigate the role of the EEG in the MBLZT process of a time-dependent quantum system, 
a commonly adopted approach is to examine the fidelity between the time-evolving wavefunction and the instantaneous eigenstates, 
denoted as ${}_{ins}\langle\psi_{n}(t)\vert\psi(t)\rangle$, which is known as the Loschmidt echo (LE).

For the time evolution governed by Eq.~(\ref{Eq:latH_driven}), 
the LE is expressed as $\mathcal{L}_{n}(t)=\langle\psi_{n}(0)\vert e^{-i\mathcal{H}_{\text{eff}}t}\vert\psi(0)\rangle$. Here, the index $n$ is used to label the energy level of the corresponding excited states.
When $n$ corresponds to the ground state, we focus on the spectrum of the LE obtained through its Fourier transformation, 
which is given by $\mathcal{L}_{0}(\omega)=\sum_{m}|\langle \phi_m|\psi_0\rangle|^2\delta(\omega-E_m)$. 
Here, $\vert\phi_m\rangle$ represents the eigenbasis of the $\mathcal{H}_{\text{eff}}$, 
and $E_{m}$ is its corresponding eigenenergy.
Detailed results for $L=80$ with different $h_z$ from $\w_0/10$ to $\w_0$ are exhibited in \cite{SM}.
To obtain a better understanding of the MBLZT, we introduce spectral entropy for the LE,
\begin{equation}
    S_{\text{LE}}=-\sum_m p_m \ln p_m,
    \label{Eq:SLE}
\end{equation}
where $p_m =|\langle \phi_m|\psi_0\rangle|^2$, and it naturally satisfies $\sum_m p_m = 1$. 
In the context of our study, the $S_{\text{LE}}$ serves as a measure of the spectral density across the entire spectrum. 

We use the TLFFA algorithm to calculate the quantities introduced above. The details of the TLFFA algorithm can be found in \cite{SM}. 
Specifically, in this algorithm, we truncate the free fermion basis to an energy of 10$J$ above the vacuum. 
For different system sizes, we ensure convergent results by increasing 
the number of states. 
For instance, for $L = 100$ the number truncation of states reaches $16384$ 
for convergence, and this also reaches our computational limit.

\begin{figure}[tp]
    \includegraphics[width=8.5cm]{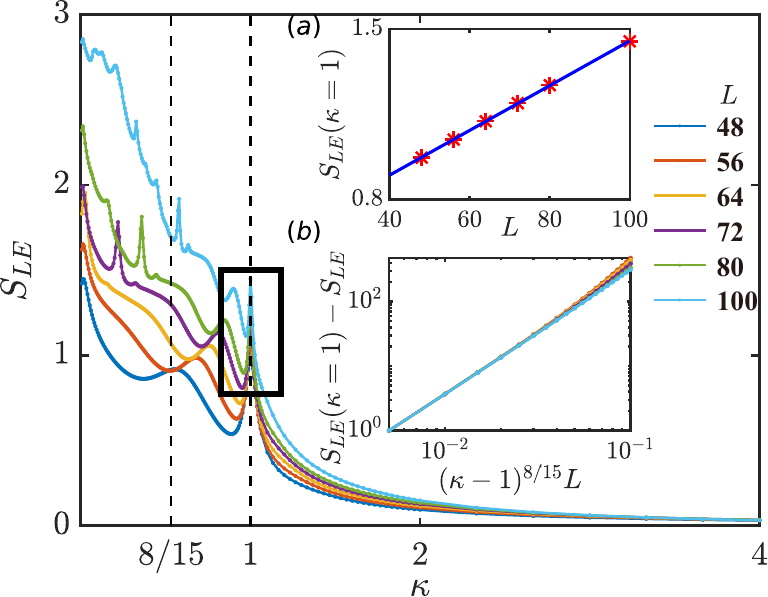}
    \caption{$S_{LE}$ $vs.$ $\kappa$ for different system sizes $L$'s. 
    The $S_{LE}$ at $\kappa=1$ linearly increases with $L$ [inset (a)], 
    along with more peaks emerging for $\kappa \lesssim 8/15$.
    When $\kappa$ approaches 1, the $S_{LE}$ collapses on the same curve for different
    $L$ [inset (b)], implying hyperscaling behavior of $S_{\text{LE}}$ w.r.t scaling
    parameter $(\kappa-1)^{8/15}L$.}    
    \label{fig:NLEw}
\end{figure}

The results are summarized in Fig.~\ref{fig:NLEw}.
When $\kappa \gg 1$, $S_{\text{LE}}$ is small, indicating weak MBLZT. As $\kappa$ decreases, 
$S_{\text{LE}}$ increases continuously, suggesting an enhancement of MBLZT. 
When $\kappa = 1$, a peak appears and increases linearly in $L$ [Fig.~\ref{fig:NLEw} inset (a)], 
which possibly implies MBLZT avalanche in thermodynamic limit.
As $\kappa$ decreases further below $8/15$ upper bound of zeros of the EEG 
between the ground state and excited states, 
it is expected that the MBLZT is further strengthened, 
as manifested in the continuously increasing $S_{\text{LE}}$. Additional peaks
appear when $\kappa < 1$, signaling further strengthened MBLZT at specific $\kappa$'s.
For $\kappa\rightarrow 0^{+}$, it is noticed that the increase of $S_{\text{LE}}$ 
becomes weaker for $L=100$ comparing to smaller $L$ [Fig.~\ref{fig:NLEw} (b)], which
possibly is due to the truncated energy cut at 10$J$ in our numerical calculation. 
A higher cut should provide better results,
yet it exceeds our computational limit.
Interestingly with $\kappa$ approaching to 1, $S_{\text{LE}}$ exhibits a hyperscaling behavior 
in terms of the scaling parameter $(\kappa-1)^{8/15}JL$ [Fig.~\ref{fig:NLEw} inset (b),
$J = 1$], implying emergence of universal physics near $\kappa = 1$. 

The entire process can also be qualitatively understood in terms of the $\mathcal{H}_{\text{eff}}$ [Eq.~(\ref{Eq:Heff})]. 
Precisely at $\kappa = 1$, the $\mathcal{H}_{\text{eff}}$ represents a quantum critical transverse field Ising chain with divergent density of states in the low energy sector, leading to a significant increase of the spectral entropy.
When $\kappa < 1$, the spin alignment in the ground state of $\mathcal{H}_{\text{eff}}$ is opposite to that of the initial ground state. This means that the initial state becomes a high-energy state with respect to the $\mathcal{H}_{\text{eff}}$, resulting in progressively stronger MBLZT as $\kappa$ decreases.

\textit{Conclusions}.---
In this article we lay down a theorem revealing the quantum geometry
in time-dependent quantum many-body systems whose low-energy can be organized by
quantum integrable field theory instantaneously.
The theorem states that the BCM entries for local field densities in 
the system 
are governed by at most two-particle excitations. 
The result provides a geometric foundation for diabatic time evolutions 
mediated by the vanishing EEG due to the QGP contribution.

We then analyze a time-dependent quantum Ising chain, 
showing that the suppression of the EEG with decreasing longitudinal field enhances the MBLZT.  
Near the critical ratio $\kappa = 1$, the hyperscaling behavior of the LE spectral entropy 
signals a sharp transition in the MBLZT.  
For $\kappa < 1$, the numerical results show persistent entropy growth 
mixing with a cascade of peaks, 
reflecting continuous enhancement of MBLZT.  
Our results indicate that the LE spectral entropy can serve as 
a powerful diagnostic tool for diabatic time evolution and its associated MBLZT.  
Yet, a full physical understanding on the analytical 
structure of the LE spectral entropy, 
such as its hyperscaling behavior, 
linear in $L$ behavior at $\kappa = 1$, and additional peaks when $\kappa < 1$, 
requires more analytical and numerical efforts, a challenge that remains open 
but worth to explore further. Meanwhile it is also worth to generalize 
the theorem in quantum integrable lattice systems, 
and extend to the BCM entries for quasi-local field densities. 

The time-dependent Hamiltonian in Eq.~(\ref{Eq:latH_driven}) is experimentally accessible 
in platforms such as Rydberg atomic arrays,~\cite{coldatom1,Aquila,xiao_2024time}
a promising setup for direct experimental verification of our theoretical predictions.

\textit{Acknowledgements}.---
We thank Zongping Gong, Qicheng Tang, Thors Hans Hansson and Gabor Tak\'{a}cs for helpful discussions.
The TCSA calculation is supported by the TCSA package in~\cite{TCSA}.
The work is sponsored by National Natural Science Foundation of China Nos. 12274288, 12450004, 
the Innovation Program for Quantum Science and Technology Grant No. 2021ZD0301900, and 
the Tsung-Dao Lee Scholarship.

\textit{Data availability}.---The data are available from the corresponding author upon reasonable request.

\bibliography{main}

\begin{thebibliography}{73}%
\makeatletter
\providecommand \@ifxundefined [1]{%
 \@ifx{#1\undefined}
}%
\providecommand \@ifnum [1]{%
 \ifnum #1\expandafter \@firstoftwo
 \else \expandafter \@secondoftwo
 \fi
}%
\providecommand \@ifx [1]{%
 \ifx #1\expandafter \@firstoftwo
 \else \expandafter \@secondoftwo
 \fi
}%
\providecommand \natexlab [1]{#1}%
\providecommand \enquote  [1]{``#1''}%
\providecommand \bibnamefont  [1]{#1}%
\providecommand \bibfnamefont [1]{#1}%
\providecommand \citenamefont [1]{#1}%
\providecommand \href@noop [0]{\@secondoftwo}%
\providecommand \href [0]{\begingroup \@sanitize@url \@href}%
\providecommand \@href[1]{\@@startlink{#1}\@@href}%
\providecommand \@@href[1]{\endgroup#1\@@endlink}%
\providecommand \@sanitize@url [0]{\catcode `\\12\catcode `\$12\catcode
  `\&12\catcode `\#12\catcode `\^12\catcode `\_12\catcode `\%12\relax}%
\providecommand \@@startlink[1]{}%
\providecommand \@@endlink[0]{}%
\providecommand \url  [0]{\begingroup\@sanitize@url \@url }%
\providecommand \@url [1]{\endgroup\@href {#1}{\urlprefix }}%
\providecommand \urlprefix  [0]{URL }%
\providecommand \Eprint [0]{\href }%
\providecommand \doibase [0]{https://doi.org/}%
\providecommand \selectlanguage [0]{\@gobble}%
\providecommand \bibinfo  [0]{\@secondoftwo}%
\providecommand \bibfield  [0]{\@secondoftwo}%
\providecommand \translation [1]{[#1]}%
\providecommand \BibitemOpen [0]{}%
\providecommand \bibitemStop [0]{}%
\providecommand \bibitemNoStop [0]{.\EOS\space}%
\providecommand \EOS [0]{\spacefactor3000\relax}%
\providecommand \BibitemShut  [1]{\csname bibitem#1\endcsname}%
\let\auto@bib@innerbib\@empty
\bibitem [{\citenamefont {Berry}(1984)}]{berryphase}%
  \BibitemOpen
  \bibfield  {author} {\bibinfo {author} {\bibfnamefont {M.~V.}\ \bibnamefont
  {Berry}},\ }\bibfield  {title} {\bibinfo {title} {{Quantal phase factors
  accompanying adiabatic changes}},\ }\href
  {https://doi.org/doi.org/10.1098/rspa.1984.0023} {\bibfield  {journal}
  {\bibinfo  {journal} {Proc. R. Soc. Lond. A}\ }\textbf {\bibinfo {volume}
  {392}},\ \bibinfo {pages} {45} (\bibinfo {year} {1984})}\BibitemShut
  {NoStop}%
\bibitem [{\citenamefont {Ahn}\ \emph {et~al.}(2020)\citenamefont {Ahn},
  \citenamefont {Guo},\ and\ \citenamefont {Nagaosa}}]{guangyu_prx}%
  \BibitemOpen
  \bibfield  {author} {\bibinfo {author} {\bibfnamefont {J.}~\bibnamefont
  {Ahn}}, \bibinfo {author} {\bibfnamefont {G.-Y.}\ \bibnamefont {Guo}},\ and\
  \bibinfo {author} {\bibfnamefont {N.}~\bibnamefont {Nagaosa}},\ }\bibfield
  {title} {\bibinfo {title} {Low-frequency divergence and quantum geometry of
  the bulk photovoltaic effect in topological semimetals},\ }\href
  {https://doi.org/10.1103/PhysRevX.10.041041} {\bibfield  {journal} {\bibinfo
  {journal} {Phys. Rev. X}\ }\textbf {\bibinfo {volume} {10}},\ \bibinfo
  {pages} {041041} (\bibinfo {year} {2020})}\BibitemShut {NoStop}%
\bibitem [{\citenamefont {Ahn}\ \emph {et~al.}(2022)\citenamefont {Ahn},
  \citenamefont {Guo}, \citenamefont {Nagaosa},\ and\ \citenamefont
  {Vishwanath}}]{guangyu_np}%
  \BibitemOpen
  \bibfield  {author} {\bibinfo {author} {\bibfnamefont {J.}~\bibnamefont
  {Ahn}}, \bibinfo {author} {\bibfnamefont {G.-Y.}\ \bibnamefont {Guo}},
  \bibinfo {author} {\bibfnamefont {N.}~\bibnamefont {Nagaosa}},\ and\ \bibinfo
  {author} {\bibfnamefont {A.}~\bibnamefont {Vishwanath}},\ }\bibfield  {title}
  {\bibinfo {title} {Riemannian geometry of resonant optical responses},\
  }\href {https://doi.org/doi.org/10.1038/s41567-021-01465-z} {\bibfield
  {journal} {\bibinfo  {journal} {Nat. Phys.}\ }\textbf {\bibinfo {volume}
  {18}},\ \bibinfo {pages} {290} (\bibinfo {year} {2022})}\BibitemShut
  {NoStop}%
\bibitem [{\citenamefont {Huang}\ \emph {et~al.}(2023)\citenamefont {Huang},
  \citenamefont {Chan},\ and\ \citenamefont {Guo}}]{gy_prb2023}%
  \BibitemOpen
  \bibfield  {author} {\bibinfo {author} {\bibfnamefont {Y.-S.}\ \bibnamefont
  {Huang}}, \bibinfo {author} {\bibfnamefont {Y.-H.}\ \bibnamefont {Chan}},\
  and\ \bibinfo {author} {\bibfnamefont {G.-Y.}\ \bibnamefont {Guo}},\
  }\bibfield  {title} {\bibinfo {title} {Large shift currents via in-gap and
  charge-neutral excitons in a monolayer and nanotubes of $\text{BN}$},\ }\href
  {https://doi.org/10.1103/PhysRevB.108.075413} {\bibfield  {journal} {\bibinfo
   {journal} {Phys. Rev. B}\ }\textbf {\bibinfo {volume} {108}},\ \bibinfo
  {pages} {075413} (\bibinfo {year} {2023})}\BibitemShut {NoStop}%
\bibitem [{\citenamefont {Takayoshi}\ \emph {et~al.}(2021)\citenamefont
  {Takayoshi}, \citenamefont {Wu},\ and\ \citenamefont {Oka}}]{da_tunneling}%
  \BibitemOpen
  \bibfield  {author} {\bibinfo {author} {\bibfnamefont {S.}~\bibnamefont
  {Takayoshi}}, \bibinfo {author} {\bibfnamefont {J.}~\bibnamefont {Wu}},\ and\
  \bibinfo {author} {\bibfnamefont {T.}~\bibnamefont {Oka}},\ }\bibfield
  {title} {\bibinfo {title} {{Nonadiabatic nonlinear optics and quantum
  geometry \-- Application to the twisted Schwinger effect}},\ }\href
  {https://doi.org/10.21468/SciPostPhys.11.4.075} {\bibfield  {journal}
  {\bibinfo  {journal} {SciPost Phys.}\ }\textbf {\bibinfo {volume} {11}},\
  \bibinfo {pages} {075} (\bibinfo {year} {2021})}\BibitemShut {NoStop}%
\bibitem [{\citenamefont {Albash}\ and\ \citenamefont
  {Lidar}(2018)}]{RMP_adiabatic}%
  \BibitemOpen
  \bibfield  {author} {\bibinfo {author} {\bibfnamefont {T.}~\bibnamefont
  {Albash}}\ and\ \bibinfo {author} {\bibfnamefont {D.~A.}\ \bibnamefont
  {Lidar}},\ }\bibfield  {title} {\bibinfo {title} {Adiabatic quantum
  computation},\ }\href {https://doi.org/10.1103/RevModPhys.90.015002}
  {\bibfield  {journal} {\bibinfo  {journal} {Rev. Mod. Phys.}\ }\textbf
  {\bibinfo {volume} {90}},\ \bibinfo {pages} {015002} (\bibinfo {year}
  {2018})}\BibitemShut {NoStop}%
\bibitem [{\citenamefont {Thouless}\ \emph {et~al.}(1982)\citenamefont
  {Thouless}, \citenamefont {Kohmoto}, \citenamefont {Nightingale},\ and\
  \citenamefont {den Nijs}}]{PhysRevLett.49.405}%
  \BibitemOpen
  \bibfield  {author} {\bibinfo {author} {\bibfnamefont {D.~J.}\ \bibnamefont
  {Thouless}}, \bibinfo {author} {\bibfnamefont {M.}~\bibnamefont {Kohmoto}},
  \bibinfo {author} {\bibfnamefont {M.~P.}\ \bibnamefont {Nightingale}},\ and\
  \bibinfo {author} {\bibfnamefont {M.}~\bibnamefont {den Nijs}},\ }\bibfield
  {title} {\bibinfo {title} {Quantized hall conductance in a two-dimensional
  periodic potential},\ }\href {https://doi.org/10.1103/PhysRevLett.49.405}
  {\bibfield  {journal} {\bibinfo  {journal} {Phys. Rev. Lett.}\ }\textbf
  {\bibinfo {volume} {49}},\ \bibinfo {pages} {405} (\bibinfo {year}
  {1982})}\BibitemShut {NoStop}%
\bibitem [{\citenamefont {Korepin}\ and\ \citenamefont
  {Wu}(1991)}]{korepin_1991}%
  \BibitemOpen
  \bibfield  {author} {\bibinfo {author} {\bibfnamefont {V.}~\bibnamefont
  {Korepin}}\ and\ \bibinfo {author} {\bibfnamefont {A.}~\bibnamefont {Wu}},\
  }\bibfield  {title} {\bibinfo {title} {Adiabatic transport properties and
  berry's phase in heisenberg-ising ring},\ }\href
  {https://doi.org/doi.org/10.1142/S0217979291000304} {\bibfield  {journal}
  {\bibinfo  {journal} {Int. J. Mod. Phys. B}\ }\textbf {\bibinfo {volume}
  {05}},\ \bibinfo {pages} {497} (\bibinfo {year} {1991})}\BibitemShut
  {NoStop}%
\bibitem [{\citenamefont {Lieb}\ and\ \citenamefont
  {Liniger}(1963)}]{LL_I_1963}%
  \BibitemOpen
  \bibfield  {author} {\bibinfo {author} {\bibfnamefont {E.~H.}\ \bibnamefont
  {Lieb}}\ and\ \bibinfo {author} {\bibfnamefont {W.}~\bibnamefont {Liniger}},\
  }\bibfield  {title} {\bibinfo {title} {Exact {Analysis} of an {Interacting}
  {Bose} {Gas}. {I}. {The} {General Solution} and the {Ground} {State}},\
  }\href {https://doi.org/10.1103/PhysRev.130.1605} {\bibfield  {journal}
  {\bibinfo  {journal} {Phys. Rev.}\ }\textbf {\bibinfo {volume} {130}},\
  \bibinfo {pages} {1605} (\bibinfo {year} {1963})}\BibitemShut {NoStop}%
\bibitem [{\citenamefont {Lieb}(1963)}]{LL_II_1963}%
  \BibitemOpen
  \bibfield  {author} {\bibinfo {author} {\bibfnamefont {E.~H.}\ \bibnamefont
  {Lieb}},\ }\bibfield  {title} {\bibinfo {title} {{Exact Analysis of an
  Interacting Bose Gas. II. The Excitation Spectrum}},\ }\href
  {https://doi.org/10.1103/PhysRev.130.1616} {\bibfield  {journal} {\bibinfo
  {journal} {Phys. Rev.}\ }\textbf {\bibinfo {volume} {130}},\ \bibinfo {pages}
  {1616} (\bibinfo {year} {1963})}\BibitemShut {NoStop}%
\bibitem [{\citenamefont {Gaudin}(1967)}]{GAUDIN1967}%
  \BibitemOpen
  \bibfield  {author} {\bibinfo {author} {\bibfnamefont {M.}~\bibnamefont
  {Gaudin}},\ }\bibfield  {title} {\bibinfo {title} {Un systeme a une dimension
  de fermions en interaction},\ }\href
  {https://doi.org/https://doi.org/10.1016/0375-9601(67)90193-4} {\bibfield
  {journal} {\bibinfo  {journal} {Phys. Lett. A}\ }\textbf {\bibinfo {volume}
  {24}},\ \bibinfo {pages} {55} (\bibinfo {year} {1967})}\BibitemShut {NoStop}%
\bibitem [{\citenamefont {Yang}(1967)}]{CNYang_1967}%
  \BibitemOpen
  \bibfield  {author} {\bibinfo {author} {\bibfnamefont {C.~N.}\ \bibnamefont
  {Yang}},\ }\bibfield  {title} {\bibinfo {title} {Some exact results for the
  many-body problem in one dimension with repulsive delta-function
  interaction},\ }\href {https://doi.org/10.1103/PhysRevLett.19.1312}
  {\bibfield  {journal} {\bibinfo  {journal} {Phys. Rev. Lett.}\ }\textbf
  {\bibinfo {volume} {19}},\ \bibinfo {pages} {1312} (\bibinfo {year}
  {1967})}\BibitemShut {NoStop}%
\bibitem [{\citenamefont {Jimbo}\ and\ \citenamefont {Miwa}(1995)}]{Jimbo1995}%
  \BibitemOpen
  \bibfield  {author} {\bibinfo {author} {\bibfnamefont {M.}~\bibnamefont
  {Jimbo}}\ and\ \bibinfo {author} {\bibfnamefont {T.}~\bibnamefont {Miwa}},\
  }\href@noop {} {\emph {\bibinfo {title} {{Algebraic Analysis of Solvable
  Lattice Models}}}}\ (\bibinfo  {publisher} {American Mathematical Society},\
  \bibinfo {year} {1995})\BibitemShut {NoStop}%
\bibitem [{\citenamefont {Guan}\ \emph {et~al.}(2013)\citenamefont {Guan},
  \citenamefont {Batchelor},\ and\ \citenamefont {Lee}}]{Guan_Fermi_2013}%
  \BibitemOpen
  \bibfield  {author} {\bibinfo {author} {\bibfnamefont {X.-W.}\ \bibnamefont
  {Guan}}, \bibinfo {author} {\bibfnamefont {M.~T.}\ \bibnamefont
  {Batchelor}},\ and\ \bibinfo {author} {\bibfnamefont {C.}~\bibnamefont
  {Lee}},\ }\bibfield  {title} {\bibinfo {title} {Fermi gases in one dimension:
  From {Bethe} ansatz to experiments},\ }\href
  {https://doi.org/10.1103/RevModPhys.85.1633} {\bibfield  {journal} {\bibinfo
  {journal} {Rev. Mod. Phys.}\ }\textbf {\bibinfo {volume} {85}},\ \bibinfo
  {pages} {1633} (\bibinfo {year} {2013})}\BibitemShut {NoStop}%
\bibitem [{\citenamefont {He}\ \emph {et~al.}(2017)\citenamefont {He},
  \citenamefont {Jiang}, \citenamefont {Yu}, \citenamefont {Lin},\ and\
  \citenamefont {Guan}}]{He_quantum_2017}%
  \BibitemOpen
  \bibfield  {author} {\bibinfo {author} {\bibfnamefont {F.}~\bibnamefont
  {He}}, \bibinfo {author} {\bibfnamefont {Y.}~\bibnamefont {Jiang}}, \bibinfo
  {author} {\bibfnamefont {Y.-C.}\ \bibnamefont {Yu}}, \bibinfo {author}
  {\bibfnamefont {H.-Q.}\ \bibnamefont {Lin}},\ and\ \bibinfo {author}
  {\bibfnamefont {X.-W.}\ \bibnamefont {Guan}},\ }\bibfield  {title} {\bibinfo
  {title} {Quantum criticality of spinons},\ }\href
  {https://doi.org/10.1103/PhysRevB.96.220401} {\bibfield  {journal} {\bibinfo
  {journal} {Phys. Rev. B}\ }\textbf {\bibinfo {volume} {96}},\ \bibinfo
  {pages} {220401} (\bibinfo {year} {2017})}\BibitemShut {NoStop}%
\bibitem [{\citenamefont {{Franchini}}(2017)}]{Franchini2017}%
  \BibitemOpen
  \bibfield  {author} {\bibinfo {author} {\bibfnamefont {F.}~\bibnamefont
  {{Franchini}}},\ }\href {https://doi.org/10.1007/978-3-319-48487-7} {\emph
  {\bibinfo {title} {{An Introduction to Integrable Techniques for
  One-Dimensional Quantum Systems}}}},\ Vol.\ \bibinfo {volume} {940}\
  (\bibinfo  {publisher} {Springer, Cham},\ \bibinfo {year} {2017})\BibitemShut
  {NoStop}%
\bibitem [{\citenamefont {Yang}\ \emph {et~al.}(2023)\citenamefont {Yang},
  \citenamefont {Wang},\ and\ \citenamefont {Wu}}]{yang_magnetic_2023}%
  \BibitemOpen
  \bibfield  {author} {\bibinfo {author} {\bibfnamefont {J.}~\bibnamefont
  {Yang}}, \bibinfo {author} {\bibfnamefont {X.}~\bibnamefont {Wang}},\ and\
  \bibinfo {author} {\bibfnamefont {J.}~\bibnamefont {Wu}},\ }\bibfield
  {title} {\bibinfo {title} {Magnetic excitations in the one-dimensional
  {Heisenberg}-{Ising} model with external fields and their experimental
  realizations},\ }\href {https://doi.org/10.1088/1751-8121/acad48} {\bibfield
  {journal} {\bibinfo  {journal} {J. Phys. A: Math. Theor.}\ }\textbf {\bibinfo
  {volume} {56}},\ \bibinfo {pages} {013001} (\bibinfo {year}
  {2023})}\BibitemShut {NoStop}%
\bibitem [{\citenamefont {Yang}\ and\ \citenamefont {Wu}(2024)}]{yjh_2024}%
  \BibitemOpen
  \bibfield  {author} {\bibinfo {author} {\bibfnamefont {J.}~\bibnamefont
  {Yang}}\ and\ \bibinfo {author} {\bibfnamefont {J.}~\bibnamefont {Wu}},\
  }\bibfield  {title} {\bibinfo {title} {Truncated string state space approach
  and its application to the nonintegrable spin-$\frac{1}{2}$ {Heisenberg}
  chain},\ }\href {https://doi.org/10.1103/PhysRevB.109.214421} {\bibfield
  {journal} {\bibinfo  {journal} {Phys. Rev. B}\ }\textbf {\bibinfo {volume}
  {109}},\ \bibinfo {pages} {214421} (\bibinfo {year} {2024})}\BibitemShut
  {NoStop}%
\bibitem [{\citenamefont {Korepin}\ \emph {et~al.}(1993)\citenamefont
  {Korepin}, \citenamefont {Bogoliubov},\ and\ \citenamefont
  {Izergin}}]{Korepin_1993}%
  \BibitemOpen
  \bibfield  {author} {\bibinfo {author} {\bibfnamefont {V.~E.}\ \bibnamefont
  {Korepin}}, \bibinfo {author} {\bibfnamefont {N.~M.}\ \bibnamefont
  {Bogoliubov}},\ and\ \bibinfo {author} {\bibfnamefont {A.~G.}\ \bibnamefont
  {Izergin}},\ }\href {https://doi.org/doi.org/10.1017/CBO9780511628832} {\emph
  {\bibinfo {title} {Quantum Inverse Scattering Method and Correlation
  Functions}}}\ (\bibinfo  {publisher} {Cambridge University Press},\ \bibinfo
  {year} {1993})\BibitemShut {NoStop}%
\bibitem [{\citenamefont {Sutherland}(2004)}]{BM_2004}%
  \BibitemOpen
  \bibfield  {author} {\bibinfo {author} {\bibfnamefont {B.}~\bibnamefont
  {Sutherland}},\ }\href {https://doi.org/10.1142/5552} {\emph {\bibinfo
  {title} {Beautiful Models}}}\ (\bibinfo  {publisher} {World Scientifc},\
  \bibinfo {year} {2004})\BibitemShut {NoStop}%
\bibitem [{\citenamefont {Smirnov}(1992)}]{smirnov_1992}%
  \BibitemOpen
  \bibfield  {author} {\bibinfo {author} {\bibfnamefont {F.~A.}\ \bibnamefont
  {Smirnov}},\ }\href {https://doi.org/10.1142/1115} {\emph {\bibinfo {title}
  {Form Factors in Completely Integrable Models of Quantum Field Theory}}}\
  (\bibinfo  {publisher} {World Scientifc},\ \bibinfo {year}
  {1992})\BibitemShut {NoStop}%
\bibitem [{\citenamefont {Delfino}\ and\ \citenamefont
  {Mussardo}(1995)}]{delfino_1995}%
  \BibitemOpen
  \bibfield  {author} {\bibinfo {author} {\bibfnamefont {G.}~\bibnamefont
  {Delfino}}\ and\ \bibinfo {author} {\bibfnamefont {G.}~\bibnamefont
  {Mussardo}},\ }\bibfield  {title} {\bibinfo {title} {The spin-spin
  correlation function in the two-dimensional ising model in a magnetic field
  at $\textit{T} = \textit{T}_c$},\ }\href
  {https://doi.org/10.1016/0550-3213(95)00464-4} {\bibfield  {journal}
  {\bibinfo  {journal} {Nucl. Phys. B}\ }\textbf {\bibinfo {volume} {455}},\
  \bibinfo {pages} {724 } (\bibinfo {year} {1995})}\BibitemShut {NoStop}%
\bibitem [{\citenamefont {H\'{o}ds\'{a}gi}\ and\ \citenamefont
  {Kormos}(2020)}]{kh2020}%
  \BibitemOpen
  \bibfield  {author} {\bibinfo {author} {\bibfnamefont {K.}~\bibnamefont
  {H\'{o}ds\'{a}gi}}\ and\ \bibinfo {author} {\bibfnamefont {M.}~\bibnamefont
  {Kormos}},\ }\bibfield  {title} {\bibinfo {title} {Kibble-zurek mechanism in
  the {Ising} {Field} {Theory}},\ }\href
  {https://doi.org/10.21468/SciPostPhys.9.4.055} {\bibfield  {journal}
  {\bibinfo  {journal} {SciPost Phys.}\ }\textbf {\bibinfo {volume} {9}},\
  \bibinfo {pages} {055} (\bibinfo {year} {2020})}\BibitemShut {NoStop}%
\bibitem [{\citenamefont {Wilczek}\ and\ \citenamefont
  {Zee}(1984)}]{PhysRevLett.52.2111}%
  \BibitemOpen
  \bibfield  {author} {\bibinfo {author} {\bibfnamefont {F.}~\bibnamefont
  {Wilczek}}\ and\ \bibinfo {author} {\bibfnamefont {A.}~\bibnamefont {Zee}},\
  }\bibfield  {title} {\bibinfo {title} {Appearance of gauge structure in
  simple dynamical systems},\ }\href
  {https://doi.org/10.1103/PhysRevLett.52.2111} {\bibfield  {journal} {\bibinfo
   {journal} {Phys. Rev. Lett.}\ }\textbf {\bibinfo {volume} {52}},\ \bibinfo
  {pages} {2111} (\bibinfo {year} {1984})}\BibitemShut {NoStop}%
\bibitem [{\citenamefont {Wu}\ \emph {et~al.}(2008)\citenamefont {Wu},
  \citenamefont {Zhao}, \citenamefont {Chen},\ and\ \citenamefont
  {Zhang}}]{da_QGP2008}%
  \BibitemOpen
  \bibfield  {author} {\bibinfo {author} {\bibfnamefont {J.-d.}\ \bibnamefont
  {Wu}}, \bibinfo {author} {\bibfnamefont {M.-s.}\ \bibnamefont {Zhao}},
  \bibinfo {author} {\bibfnamefont {J.-l.}\ \bibnamefont {Chen}},\ and\
  \bibinfo {author} {\bibfnamefont {Y.-d.}\ \bibnamefont {Zhang}},\ }\bibfield
  {title} {\bibinfo {title} {Adiabatic condition and quantum geometric
  potential},\ }\href {https://doi.org/10.1103/PhysRevA.77.062114} {\bibfield
  {journal} {\bibinfo  {journal} {Phys. Rev. A}\ }\textbf {\bibinfo {volume}
  {77}},\ \bibinfo {pages} {062114} (\bibinfo {year} {2008})}\BibitemShut
  {NoStop}%
\bibitem [{\citenamefont {Xu}\ \emph {et~al.}(2018)\citenamefont {Xu},
  \citenamefont {Wu},\ and\ \citenamefont {Wu}}]{da_QGP2018}%
  \BibitemOpen
  \bibfield  {author} {\bibinfo {author} {\bibfnamefont {C.}~\bibnamefont
  {Xu}}, \bibinfo {author} {\bibfnamefont {J.}~\bibnamefont {Wu}},\ and\
  \bibinfo {author} {\bibfnamefont {C.}~\bibnamefont {Wu}},\ }\bibfield
  {title} {\bibinfo {title} {Quantized interlevel character in quantum
  systems},\ }\href {https://doi.org/10.1103/PhysRevA.97.032124} {\bibfield
  {journal} {\bibinfo  {journal} {Phys. Rev. A}\ }\textbf {\bibinfo {volume}
  {97}},\ \bibinfo {pages} {032124} (\bibinfo {year} {2018})}\BibitemShut
  {NoStop}%
\bibitem [{\citenamefont {Delfino}(2014)}]{delfino_2014}%
  \BibitemOpen
  \bibfield  {author} {\bibinfo {author} {\bibfnamefont {G.}~\bibnamefont
  {Delfino}},\ }\bibfield  {title} {\bibinfo {title} {Quantum quenches with
  integrable pre-quench dynamics},\ }\href
  {https://doi.org/10.1088/1751-8113/47/40/402001} {\bibfield  {journal}
  {\bibinfo  {journal} {J. Phys A: Math. Theor.}\ }\textbf {\bibinfo {volume}
  {47}},\ \bibinfo {pages} {402001} (\bibinfo {year} {2014})}\BibitemShut
  {NoStop}%
\bibitem [{\citenamefont {Delfino}\ and\ \citenamefont
  {Sorba}(2024)}]{delfino_2024}%
  \BibitemOpen
  \bibfield  {author} {\bibinfo {author} {\bibfnamefont {G.}~\bibnamefont
  {Delfino}}\ and\ \bibinfo {author} {\bibfnamefont {M.}~\bibnamefont
  {Sorba}},\ }\bibfield  {title} {\bibinfo {title} {On unitary time evolution
  out of equilibrium},\ }\href
  {https://doi.org/doi.org/10.1016/j.nuclphysb.2024.116587} {\bibfield
  {journal} {\bibinfo  {journal} {Nucl. Phys. B}\ }\textbf {\bibinfo {volume}
  {1005}},\ \bibinfo {pages} {116587 } (\bibinfo {year} {2024})}\BibitemShut
  {NoStop}%
\bibitem [{\citenamefont {Caux}\ and\ \citenamefont
  {Essler}(2013)}]{PhysRevLett.110.257203}%
  \BibitemOpen
  \bibfield  {author} {\bibinfo {author} {\bibfnamefont {J.-S.}\ \bibnamefont
  {Caux}}\ and\ \bibinfo {author} {\bibfnamefont {F.~H.~L.}\ \bibnamefont
  {Essler}},\ }\bibfield  {title} {\bibinfo {title} {Time evolution of local
  observables after quenching to an integrable model},\ }\href
  {https://doi.org/10.1103/PhysRevLett.110.257203} {\bibfield  {journal}
  {\bibinfo  {journal} {Phys. Rev. Lett.}\ }\textbf {\bibinfo {volume} {110}},\
  \bibinfo {pages} {257203} (\bibinfo {year} {2013})}\BibitemShut {NoStop}%
\bibitem [{\citenamefont {Calabrese}\ \emph {et~al.}(2011)\citenamefont
  {Calabrese}, \citenamefont {Essler},\ and\ \citenamefont
  {Fagotti}}]{CalabreseEsslerFagotti}%
  \BibitemOpen
  \bibfield  {author} {\bibinfo {author} {\bibfnamefont {P.}~\bibnamefont
  {Calabrese}}, \bibinfo {author} {\bibfnamefont {F.~H.~L.}\ \bibnamefont
  {Essler}},\ and\ \bibinfo {author} {\bibfnamefont {M.}~\bibnamefont
  {Fagotti}},\ }\bibfield  {title} {\bibinfo {title} {Quantum quench in the
  transverse-field {Ising} chain},\ }\href
  {https://doi.org/10.1103/PhysRevLett.106.227203} {\bibfield  {journal}
  {\bibinfo  {journal} {Phys. Rev. Lett.}\ }\textbf {\bibinfo {volume} {106}},\
  \bibinfo {pages} {227203} (\bibinfo {year} {2011})}\BibitemShut {NoStop}%
\bibitem [{\citenamefont {Kormos}\ \emph {et~al.}(2017)\citenamefont {Kormos},
  \citenamefont {Collura}, \citenamefont {Tak\'{a}cs},\ and\ \citenamefont
  {Calabrese}}]{marton2016}%
  \BibitemOpen
  \bibfield  {author} {\bibinfo {author} {\bibfnamefont {M.}~\bibnamefont
  {Kormos}}, \bibinfo {author} {\bibfnamefont {M.}~\bibnamefont {Collura}},
  \bibinfo {author} {\bibfnamefont {G.}~\bibnamefont {Tak\'{a}cs}},\ and\
  \bibinfo {author} {\bibfnamefont {P.}~\bibnamefont {Calabrese}},\ }\bibfield
  {title} {\bibinfo {title} {Real-time confinement following a quantum quench
  to a non-integrable model},\ }\href {https://doi.org/10.1038/nphys3934}
  {\bibfield  {journal} {\bibinfo  {journal} {Nat. Phys.}\ }\textbf {\bibinfo
  {volume} {13}},\ \bibinfo {pages} {246} (\bibinfo {year} {2017})}\BibitemShut
  {NoStop}%
\bibitem [{\citenamefont {Delfino}\ and\ \citenamefont
  {Viti}(2017)}]{Delfino_2017}%
  \BibitemOpen
  \bibfield  {author} {\bibinfo {author} {\bibfnamefont {G.}~\bibnamefont
  {Delfino}}\ and\ \bibinfo {author} {\bibfnamefont {J.}~\bibnamefont {Viti}},\
  }\bibfield  {title} {\bibinfo {title} {On the theory of quantum quenches in
  near-critical systems},\ }\href {https://doi.org/10.1088/1751-8121/aa5660}
  {\bibfield  {journal} {\bibinfo  {journal} {J. Phys A: Math. Theor.}\
  }\textbf {\bibinfo {volume} {50}},\ \bibinfo {pages} {084004} (\bibinfo
  {year} {2017})}\BibitemShut {NoStop}%
\bibitem [{\citenamefont {H\'{o}ds\'{a}gi}\ \emph {et~al.}(2018)\citenamefont
  {H\'{o}ds\'{a}gi}, \citenamefont {Kormos},\ and\ \citenamefont
  {Tak\'{a}cs}}]{kh_2018}%
  \BibitemOpen
  \bibfield  {author} {\bibinfo {author} {\bibfnamefont {K.}~\bibnamefont
  {H\'{o}ds\'{a}gi}}, \bibinfo {author} {\bibfnamefont {M.}~\bibnamefont
  {Kormos}},\ and\ \bibinfo {author} {\bibfnamefont {G.}~\bibnamefont
  {Tak\'{a}cs}},\ }\bibfield  {title} {\bibinfo {title} {Quench dynamics of the
  {Ising} field theory in a magnetic field},\ }\href
  {https://doi.org/10.21468/SciPostPhys.5.3.027} {\bibfield  {journal}
  {\bibinfo  {journal} {SciPost Phys.}\ }\textbf {\bibinfo {volume} {5}},\
  \bibinfo {pages} {027} (\bibinfo {year} {2018})}\BibitemShut {NoStop}%
\bibitem [{\citenamefont {Granet}\ \emph {et~al.}(2020)\citenamefont {Granet},
  \citenamefont {Fagotti},\ and\ \citenamefont {Essler}}]{Etienne2020}%
  \BibitemOpen
  \bibfield  {author} {\bibinfo {author} {\bibfnamefont {E.}~\bibnamefont
  {Granet}}, \bibinfo {author} {\bibfnamefont {M.}~\bibnamefont {Fagotti}},\
  and\ \bibinfo {author} {\bibfnamefont {F.~H.~L.}\ \bibnamefont {Essler}},\
  }\bibfield  {title} {\bibinfo {title} {{Finite temperature and quench
  dynamics in the Transverse Field Ising Model from form factor expansions}},\
  }\href {https://doi.org/10.21468/SciPostPhys.9.3.033} {\bibfield  {journal}
  {\bibinfo  {journal} {SciPost Phys.}\ }\textbf {\bibinfo {volume} {9}},\
  \bibinfo {pages} {033} (\bibinfo {year} {2020})}\BibitemShut {NoStop}%
\bibitem [{\citenamefont {Vanicat}\ \emph {et~al.}(2018)\citenamefont
  {Vanicat}, \citenamefont {Zadnik},\ and\ \citenamefont
  {Prosen}}]{PhysRevLett.121.030606}%
  \BibitemOpen
  \bibfield  {author} {\bibinfo {author} {\bibfnamefont {M.}~\bibnamefont
  {Vanicat}}, \bibinfo {author} {\bibfnamefont {L.}~\bibnamefont {Zadnik}},\
  and\ \bibinfo {author} {\bibfnamefont {T.~c.~v.}\ \bibnamefont {Prosen}},\
  }\bibfield  {title} {\bibinfo {title} {Integrable trotterization: Local
  conservation laws and boundary driving},\ }\href
  {https://doi.org/10.1103/PhysRevLett.121.030606} {\bibfield  {journal}
  {\bibinfo  {journal} {Phys. Rev. Lett.}\ }\textbf {\bibinfo {volume} {121}},\
  \bibinfo {pages} {030606} (\bibinfo {year} {2018})}\BibitemShut {NoStop}%
\bibitem [{\citenamefont {Vernier}\ \emph {et~al.}(2023)\citenamefont
  {Vernier}, \citenamefont {Bertini}, \citenamefont {Giudici},\ and\
  \citenamefont {Piroli}}]{PhysRevLett.130.260401}%
  \BibitemOpen
  \bibfield  {author} {\bibinfo {author} {\bibfnamefont {E.}~\bibnamefont
  {Vernier}}, \bibinfo {author} {\bibfnamefont {B.}~\bibnamefont {Bertini}},
  \bibinfo {author} {\bibfnamefont {G.}~\bibnamefont {Giudici}},\ and\ \bibinfo
  {author} {\bibfnamefont {L.}~\bibnamefont {Piroli}},\ }\bibfield  {title}
  {\bibinfo {title} {Integrable digital quantum simulation: Generalized gibbs
  ensembles and trotter transitions},\ }\href
  {https://doi.org/10.1103/PhysRevLett.130.260401} {\bibfield  {journal}
  {\bibinfo  {journal} {Phys. Rev. Lett.}\ }\textbf {\bibinfo {volume} {130}},\
  \bibinfo {pages} {260401} (\bibinfo {year} {2023})}\BibitemShut {NoStop}%
\bibitem [{\citenamefont {Hutsalyuk}\ \emph {et~al.}(2025)\citenamefont
  {Hutsalyuk}, \citenamefont {Jiang}, \citenamefont {Pozsgay}, \citenamefont
  {Xu},\ and\ \citenamefont {Zhang}}]{yunfeng_2024}%
  \BibitemOpen
  \bibfield  {author} {\bibinfo {author} {\bibfnamefont {A.}~\bibnamefont
  {Hutsalyuk}}, \bibinfo {author} {\bibfnamefont {Y.}~\bibnamefont {Jiang}},
  \bibinfo {author} {\bibfnamefont {B.}~\bibnamefont {Pozsgay}}, \bibinfo
  {author} {\bibfnamefont {H.}~\bibnamefont {Xu}},\ and\ \bibinfo {author}
  {\bibfnamefont {Y.}~\bibnamefont {Zhang}},\ }\bibfield  {title} {\bibinfo
  {title} {Exact spin correlators of integrable quantum circuits from algebraic
  geometry},\ }\href {https://doi.org/10.21468/SciPostPhys.19.1.003} {\bibfield
   {journal} {\bibinfo  {journal} {SciPost Phys.}\ }\textbf {\bibinfo {volume}
  {19}},\ \bibinfo {pages} {003} (\bibinfo {year} {2025})}\BibitemShut
  {NoStop}%
\bibitem [{\citenamefont {Wang}\ \emph
  {et~al.}(2024{\natexlab{a}})\citenamefont {Wang}, \citenamefont {Oshikawa},
  \citenamefont {Kormos},\ and\ \citenamefont {Wu}}]{xiao_2024time}%
  \BibitemOpen
  \bibfield  {author} {\bibinfo {author} {\bibfnamefont {X.}~\bibnamefont
  {Wang}}, \bibinfo {author} {\bibfnamefont {M.}~\bibnamefont {Oshikawa}},
  \bibinfo {author} {\bibfnamefont {M.}~\bibnamefont {Kormos}},\ and\ \bibinfo
  {author} {\bibfnamefont {J.}~\bibnamefont {Wu}},\ }\bibfield  {title}
  {\bibinfo {title} {Magnetization oscillations in a periodically driven
  transverse field {Ising} chain},\ }\href
  {https://doi.org/10.1103/PhysRevB.110.195101} {\bibfield  {journal} {\bibinfo
   {journal} {Phys. Rev. B}\ }\textbf {\bibinfo {volume} {110}},\ \bibinfo
  {pages} {195101} (\bibinfo {year} {2024}{\natexlab{a}})}\BibitemShut
  {NoStop}%
\bibitem [{\citenamefont {Di~Salvo}\ and\ \citenamefont
  {Schuricht}(2025)}]{DiSalvo}%
  \BibitemOpen
  \bibfield  {author} {\bibinfo {author} {\bibfnamefont {E.}~\bibnamefont
  {Di~Salvo}}\ and\ \bibinfo {author} {\bibfnamefont {D.}~\bibnamefont
  {Schuricht}},\ }\bibfield  {title} {\bibinfo {title} {Relaxation dynamics of
  integrable field theories after a global quantum quench},\ }\href
  {https://doi.org/10.1088/1742-5468/ad9f4e} {\bibfield  {journal} {\bibinfo
  {journal} {J. Stat. Mech}\ ,\ \bibinfo {pages} {013103}} (\bibinfo {year}
  {2025})}\BibitemShut {NoStop}%
\bibitem [{\citenamefont {Wiegmann}(1992)}]{Paul_1992}%
  \BibitemOpen
  \bibfield  {author} {\bibinfo {author} {\bibfnamefont {P.}~\bibnamefont
  {Wiegmann}},\ }\bibfield  {title} {\bibinfo {title} {Topological
  superconductivity},\ }\href {https://doi.org/10.1143/PTPS.107.243} {\bibfield
   {journal} {\bibinfo  {journal} {Prog. Theor. Phys. Supp.}\ }\textbf
  {\bibinfo {volume} {107}},\ \bibinfo {pages} {243} (\bibinfo {year}
  {1992})}\BibitemShut {NoStop}%
\bibitem [{\citenamefont {Robinson}\ \emph {et~al.}(2021)\citenamefont
  {Robinson}, \citenamefont {Castillo},\ and\ \citenamefont
  {Guzm\'{a}n-Gonz\'{a}lez}}]{Robinson}%
  \BibitemOpen
  \bibfield  {author} {\bibinfo {author} {\bibfnamefont {N.~J.}\ \bibnamefont
  {Robinson}}, \bibinfo {author} {\bibfnamefont {I.~P.}\ \bibnamefont
  {Castillo}},\ and\ \bibinfo {author} {\bibfnamefont {E.}~\bibnamefont
  {Guzm\'{a}n-Gonz\'{a}lez}},\ }\bibfield  {title} {\bibinfo {title} {Quantum
  quench in a driven {Ising} chain},\ }\href
  {https://doi.org/10.1103/PhysRevB.103.L140407} {\bibfield  {journal}
  {\bibinfo  {journal} {Phys. Rev. B}\ }\textbf {\bibinfo {volume} {103}},\
  \bibinfo {pages} {L140407} (\bibinfo {year} {2021})}\BibitemShut {NoStop}%
\bibitem [{\citenamefont {Du}\ \emph {et~al.}(2008)\citenamefont {Du},
  \citenamefont {Hu}, \citenamefont {Wang}, \citenamefont {Wu}, \citenamefont
  {Zhao},\ and\ \citenamefont {Suter}}]{DU_2008}%
  \BibitemOpen
  \bibfield  {author} {\bibinfo {author} {\bibfnamefont {J.}~\bibnamefont
  {Du}}, \bibinfo {author} {\bibfnamefont {L.}~\bibnamefont {Hu}}, \bibinfo
  {author} {\bibfnamefont {Y.}~\bibnamefont {Wang}}, \bibinfo {author}
  {\bibfnamefont {J.}~\bibnamefont {Wu}}, \bibinfo {author} {\bibfnamefont
  {M.}~\bibnamefont {Zhao}},\ and\ \bibinfo {author} {\bibfnamefont
  {D.}~\bibnamefont {Suter}},\ }\bibfield  {title} {\bibinfo {title}
  {Experimental study of the validity of quantitative conditions in the quantum
  adiabatic theorem},\ }\href {https://doi.org/10.1103/PhysRevLett.101.060403}
  {\bibfield  {journal} {\bibinfo  {journal} {Phys. Rev. Lett.}\ }\textbf
  {\bibinfo {volume} {101}},\ \bibinfo {pages} {060403} (\bibinfo {year}
  {2008})}\BibitemShut {NoStop}%
\bibitem [{SM()}]{SM}%
  \BibitemOpen
  \href@noop {} {}\bibinfo {howpublished} {See Supplemental Material which
  includes
  Refs.~\cite{Pfeuty,sachdev_2011,zam_1989,zam2003,Mussardo:2010mgq,Eden:1966dnq,ZAMOLODCHIKOV1979253,Zam_1989_IFT,Dorey_1996gd,CAcerbi_1997,KAROWSKI1979244,city961,Babujian:1998uw,Babujian:2001wp,DELFINO1996327,DELFINO1996450},
  for: 1. A brief introduction to the finite-size formalism of integrable
  quantum field theory and the derivation of the theorem in the main text; 2.
  Details of the derivations for the instantaneous energy levels, eigenstates,
  QGP, and EEG of the periodically driven quantum Ising chain; 3. Technical
  details of the TLFFA algorithm and an illustration of an LE
  spectrum.}\BibitemShut {Stop}%
\bibitem [{\citenamefont {Iorgov}\ \emph {et~al.}(2011)\citenamefont {Iorgov},
  \citenamefont {Shadura},\ and\ \citenamefont {Tykhyy}}]{Iorgov2011}%
  \BibitemOpen
  \bibfield  {author} {\bibinfo {author} {\bibfnamefont {N.}~\bibnamefont
  {Iorgov}}, \bibinfo {author} {\bibfnamefont {V.}~\bibnamefont {Shadura}},\
  and\ \bibinfo {author} {\bibfnamefont {Y.}~\bibnamefont {Tykhyy}},\
  }\bibfield  {title} {\bibinfo {title} {Spin operator matrix elements in the
  quantum {Ising} chain: fermion approach},\ }\href
  {https://doi.org/10.1088/1742-5468/2011/02/P02028} {\bibfield  {journal}
  {\bibinfo  {journal} {J. Stat. Mech}\ ,\ \bibinfo {pages} {P02028}} (\bibinfo
  {year} {2011})}\BibitemShut {NoStop}%
\bibitem [{\citenamefont {Yurov}\ and\ \citenamefont
  {Zamolodchikov}(1990)}]{zam_1990}%
  \BibitemOpen
  \bibfield  {author} {\bibinfo {author} {\bibfnamefont {V.}~\bibnamefont
  {Yurov}}\ and\ \bibinfo {author} {\bibfnamefont {A.}~\bibnamefont
  {Zamolodchikov}},\ }\bibfield  {title} {\bibinfo {title}
  {Truncated-fermionic-space approach to the critical 2d ising model with
  magnetic field},\ }\href {https://doi.org/doi.org/10.1142/S0217751X91002161}
  {\bibfield  {journal} {\bibinfo  {journal} {Int. J. Mod. Phys. A}\ }\textbf
  {\bibinfo {volume} {06}},\ \bibinfo {pages} {4557} (\bibinfo {year}
  {1990})}\BibitemShut {NoStop}%
\bibitem [{\citenamefont {Albert}\ \emph {et~al.}(2024)\citenamefont {Albert},
  \citenamefont {Zhang},\ and\ \citenamefont {Tu}}]{Tu2410}%
  \BibitemOpen
  \bibfield  {author} {\bibinfo {author} {\bibfnamefont {N.}~\bibnamefont
  {Albert}}, \bibinfo {author} {\bibfnamefont {Y.}~\bibnamefont {Zhang}},\ and\
  \bibinfo {author} {\bibfnamefont {H.-H.}\ \bibnamefont {Tu}},\ }\bibfield
  {title} {\bibinfo {title} {Truncated gaussian basis approach for simulating
  many-body dynamics},\ }\href {https://arxiv.org/abs/2410.04204} {\bibfield
  {journal} {\bibinfo  {journal} {arXiv:2410.04204}\ } (\bibinfo {year}
  {2024})}\BibitemShut {NoStop}%
\bibitem [{\citenamefont {Chen}\ \emph {et~al.}(2022)\citenamefont {Chen},
  \citenamefont {Fitzpatrick}, \citenamefont {Katz},\ and\ \citenamefont
  {Xin}}]{TCSA}%
  \BibitemOpen
  \bibfield  {author} {\bibinfo {author} {\bibfnamefont {H.}~\bibnamefont
  {Chen}}, \bibinfo {author} {\bibfnamefont {A.~L.}\ \bibnamefont
  {Fitzpatrick}}, \bibinfo {author} {\bibfnamefont {E.}~\bibnamefont {Katz}},\
  and\ \bibinfo {author} {\bibfnamefont {Y.}~\bibnamefont {Xin}},\ }\bibfield
  {title} {\bibinfo {title} {Giving hamiltonian truncation a boost},\ }\href
  {https://arxiv.org/abs/2207.01659} {\bibfield  {journal} {\bibinfo  {journal}
  {arXiv:2207.01659}\ } (\bibinfo {year} {2022})}\BibitemShut {NoStop}%
\bibitem [{\citenamefont {Zhang}(2010)}]{yongdezhang}%
  \BibitemOpen
  \bibfield  {author} {\bibinfo {author} {\bibfnamefont {Y.-d.}\ \bibnamefont
  {Zhang}},\ }\href {https://books.google.com/books?id=UX6KoAEACAAJ} {\emph
  {\bibinfo {title} {Advanced Quantum Mechanics}}}\ (\bibinfo  {publisher}
  {Science Press},\ \bibinfo {address} {Beijing, China},\ \bibinfo {year}
  {2010})\BibitemShut {NoStop}%
\bibitem [{\citenamefont {Pozsgay}\ and\ \citenamefont
  {Tak\'{a}cs}(2008{\natexlab{a}})}]{POZSGAY2008167}%
  \BibitemOpen
  \bibfield  {author} {\bibinfo {author} {\bibfnamefont {B.}~\bibnamefont
  {Pozsgay}}\ and\ \bibinfo {author} {\bibfnamefont {G.}~\bibnamefont
  {Tak\'{a}cs}},\ }\bibfield  {title} {\bibinfo {title} {Form factors in finite
  volume {I}: Form factor bootstrap and truncated conformal space},\ }\href
  {https://doi.org/10.1016/j.nuclphysb.2007.06.027} {\bibfield  {journal}
  {\bibinfo  {journal} {Nucl. Phys. B}\ }\textbf {\bibinfo {volume} {788}},\
  \bibinfo {pages} {167} (\bibinfo {year} {2008}{\natexlab{a}})}\BibitemShut
  {NoStop}%
\bibitem [{\citenamefont {Pozsgay}\ and\ \citenamefont
  {Tak\'{a}cs}(2008{\natexlab{b}})}]{POZSGAY2008209}%
  \BibitemOpen
  \bibfield  {author} {\bibinfo {author} {\bibfnamefont {B.}~\bibnamefont
  {Pozsgay}}\ and\ \bibinfo {author} {\bibfnamefont {G.}~\bibnamefont
  {Tak\'{a}cs}},\ }\bibfield  {title} {\bibinfo {title} {Form factors in finite
  volume {II}: Disconnected terms and finite temperature correlators},\ }\href
  {https://doi.org/10.1016/j.nuclphysb.2007.07.008} {\bibfield  {journal}
  {\bibinfo  {journal} {Nucl. Phys. B}\ }\textbf {\bibinfo {volume} {788}},\
  \bibinfo {pages} {209} (\bibinfo {year} {2008}{\natexlab{b}})}\BibitemShut
  {NoStop}%
\bibitem [{\citenamefont {H\'{o}ds\'{a}gi}\ \emph {et~al.}(2019)\citenamefont
  {H\'{o}ds\'{a}gi}, \citenamefont {Kormos},\ and\ \citenamefont
  {Tak\'{a}cs}}]{H_ds_gi_2019}%
  \BibitemOpen
  \bibfield  {author} {\bibinfo {author} {\bibfnamefont {K.}~\bibnamefont
  {H\'{o}ds\'{a}gi}}, \bibinfo {author} {\bibfnamefont {M.}~\bibnamefont
  {Kormos}},\ and\ \bibinfo {author} {\bibfnamefont {G.}~\bibnamefont
  {Tak\'{a}cs}},\ }\bibfield  {title} {\bibinfo {title} {Perturbative
  post-quench overlaps in quantum field theory},\ }\href
  {https://link.springer.com/article/10.1007/JHEP08(2019)047} {\bibfield
  {journal} {\bibinfo  {journal} {J. High Energy Phys.}\ }\bibinfo  {number} {
  (8)}}\BibitemShut {NoStop}%
\bibitem [{\citenamefont {L\"{u}scher}(1986)}]{mass}%
  \BibitemOpen
\bibfield  {number} {  }\bibfield  {author} {\bibinfo {author} {\bibfnamefont
  {M.}~\bibnamefont {L\"{u}scher}},\ }\bibfield  {title} {\bibinfo {title}
  {Volume dependence of the energy spectrum in massive quantum field theories,
  {I}. stable particle states},\ }\href
  {https://doi.org/https://doi.org/10.1007/BF01211589} {\bibfield  {journal}
  {\bibinfo  {journal} {Commun. Math. Phys.}\ }\textbf {\bibinfo {volume}
  {104}},\ \bibinfo {pages} {177} (\bibinfo {year} {1986})}\BibitemShut
  {NoStop}%
\bibitem [{\citenamefont {Fateev}(1994)}]{fateev}%
  \BibitemOpen
  \bibfield  {author} {\bibinfo {author} {\bibfnamefont {V.}~\bibnamefont
  {Fateev}},\ }\bibfield  {title} {\bibinfo {title} {The exact relations
  between the coupling constants and the masses of particles for the integrable
  perturbed conformal field theories},\ }\href
  {https://doi.org/10.1016/0370-2693(94)00078-6} {\bibfield  {journal}
  {\bibinfo  {journal} {Phys. Lett. B}\ }\textbf {\bibinfo {volume} {324}},\
  \bibinfo {pages} {45} (\bibinfo {year} {1994})}\BibitemShut {NoStop}%
\bibitem [{\citenamefont {Wang}\ \emph {et~al.}(2021)\citenamefont {Wang},
  \citenamefont {Zou}, \citenamefont {H\'ods\'agi}, \citenamefont {Kormos},
  \citenamefont {Tak\'acs},\ and\ \citenamefont {Wu}}]{xiao_2021}%
  \BibitemOpen
  \bibfield  {author} {\bibinfo {author} {\bibfnamefont {X.}~\bibnamefont
  {Wang}}, \bibinfo {author} {\bibfnamefont {H.}~\bibnamefont {Zou}}, \bibinfo
  {author} {\bibfnamefont {K.}~\bibnamefont {H\'ods\'agi}}, \bibinfo {author}
  {\bibfnamefont {M.}~\bibnamefont {Kormos}}, \bibinfo {author} {\bibfnamefont
  {G.}~\bibnamefont {Tak\'acs}},\ and\ \bibinfo {author} {\bibfnamefont
  {J.}~\bibnamefont {Wu}},\ }\bibfield  {title} {\bibinfo {title} {Cascade of
  singularities in the spin dynamics of a perturbed quantum critical {Ising}
  chain},\ }\href {https://doi.org/10.1103/PhysRevB.103.235117} {\bibfield
  {journal} {\bibinfo  {journal} {Phys. Rev. B}\ }\textbf {\bibinfo {volume}
  {103}},\ \bibinfo {pages} {235117} (\bibinfo {year} {2021})}\BibitemShut
  {NoStop}%
\bibitem [{\citenamefont {Wang}\ \emph
  {et~al.}(2024{\natexlab{b}})\citenamefont {Wang}, \citenamefont {Puzniak},
  \citenamefont {Schmalzl}, \citenamefont {Balz}, \citenamefont {Matsuda},
  \citenamefont {Okutani}, \citenamefont {Hagiwara}, \citenamefont {Ma},
  \citenamefont {Wu},\ and\ \citenamefont {Lake}}]{xiao_2023}%
  \BibitemOpen
  \bibfield  {author} {\bibinfo {author} {\bibfnamefont {X.}~\bibnamefont
  {Wang}}, \bibinfo {author} {\bibfnamefont {K.}~\bibnamefont {Puzniak}},
  \bibinfo {author} {\bibfnamefont {K.}~\bibnamefont {Schmalzl}}, \bibinfo
  {author} {\bibfnamefont {C.}~\bibnamefont {Balz}}, \bibinfo {author}
  {\bibfnamefont {M.}~\bibnamefont {Matsuda}}, \bibinfo {author} {\bibfnamefont
  {A.}~\bibnamefont {Okutani}}, \bibinfo {author} {\bibfnamefont
  {M.}~\bibnamefont {Hagiwara}}, \bibinfo {author} {\bibfnamefont
  {J.}~\bibnamefont {Ma}}, \bibinfo {author} {\bibfnamefont {J.}~\bibnamefont
  {Wu}},\ and\ \bibinfo {author} {\bibfnamefont {B.}~\bibnamefont {Lake}},\
  }\bibfield  {title} {\bibinfo {title} {Spin dynamics of the ${E}_{8}$
  particles},\ }\href
  {https://doi.org/https://doi.org/10.1016/j.scib.2024.07.040} {\bibfield
  {journal} {\bibinfo  {journal} {Sci. Bull.}\ }\textbf {\bibinfo {volume}
  {69}},\ \bibinfo {pages} {2974} (\bibinfo {year}
  {2024}{\natexlab{b}})}\BibitemShut {NoStop}%
\bibitem [{\citenamefont {Monroe}\ \emph {et~al.}(2021)\citenamefont {Monroe},
  \citenamefont {Campbell}, \citenamefont {Duan}, \citenamefont {Gong},
  \citenamefont {Gorshkov}, \citenamefont {Hess}, \citenamefont {Islam},
  \citenamefont {Kim}, \citenamefont {Linke}, \citenamefont {Pagano},
  \citenamefont {Richerme}, \citenamefont {Senko},\ and\ \citenamefont
  {Yao}}]{coldatom1}%
  \BibitemOpen
  \bibfield  {author} {\bibinfo {author} {\bibfnamefont {C.}~\bibnamefont
  {Monroe}}, \bibinfo {author} {\bibfnamefont {W.~C.}\ \bibnamefont
  {Campbell}}, \bibinfo {author} {\bibfnamefont {L.-M.}\ \bibnamefont {Duan}},
  \bibinfo {author} {\bibfnamefont {Z.-X.}\ \bibnamefont {Gong}}, \bibinfo
  {author} {\bibfnamefont {A.~V.}\ \bibnamefont {Gorshkov}}, \bibinfo {author}
  {\bibfnamefont {P.~W.}\ \bibnamefont {Hess}}, \bibinfo {author}
  {\bibfnamefont {R.}~\bibnamefont {Islam}}, \bibinfo {author} {\bibfnamefont
  {K.}~\bibnamefont {Kim}}, \bibinfo {author} {\bibfnamefont {N.~M.}\
  \bibnamefont {Linke}}, \bibinfo {author} {\bibfnamefont {G.}~\bibnamefont
  {Pagano}}, \bibinfo {author} {\bibfnamefont {P.}~\bibnamefont {Richerme}},
  \bibinfo {author} {\bibfnamefont {C.}~\bibnamefont {Senko}},\ and\ \bibinfo
  {author} {\bibfnamefont {N.~Y.}\ \bibnamefont {Yao}},\ }\bibfield  {title}
  {\bibinfo {title} {Programmable quantum simulations of spin systems with
  trapped ions},\ }\href {https://doi.org/10.1103/RevModPhys.93.025001}
  {\bibfield  {journal} {\bibinfo  {journal} {Rev. Mod. Phys.}\ }\textbf
  {\bibinfo {volume} {93}},\ \bibinfo {pages} {025001} (\bibinfo {year}
  {2021})}\BibitemShut {NoStop}%
\bibitem [{\citenamefont {Wurtz}\ \emph {et~al.}(2023)\citenamefont {Wurtz},
  \citenamefont {Bylinskii}, \citenamefont {Braverman}, \citenamefont
  {Amato-Grill}, \citenamefont {Cantu}, \citenamefont {Huber}, \citenamefont
  {Lukin}, \citenamefont {Liu}, \citenamefont {Weinberg}, \citenamefont {Long},
  \citenamefont {Wang}, \citenamefont {Gemelke},\ and\ \citenamefont
  {Keesling}}]{Aquila}%
  \BibitemOpen
  \bibfield  {author} {\bibinfo {author} {\bibfnamefont {J.}~\bibnamefont
  {Wurtz}}, \bibinfo {author} {\bibfnamefont {A.}~\bibnamefont {Bylinskii}},
  \bibinfo {author} {\bibfnamefont {B.}~\bibnamefont {Braverman}}, \bibinfo
  {author} {\bibfnamefont {J.}~\bibnamefont {Amato-Grill}}, \bibinfo {author}
  {\bibfnamefont {S.~H.}\ \bibnamefont {Cantu}}, \bibinfo {author}
  {\bibfnamefont {F.}~\bibnamefont {Huber}}, \bibinfo {author} {\bibfnamefont
  {A.}~\bibnamefont {Lukin}}, \bibinfo {author} {\bibfnamefont
  {F.}~\bibnamefont {Liu}}, \bibinfo {author} {\bibfnamefont {P.}~\bibnamefont
  {Weinberg}}, \bibinfo {author} {\bibfnamefont {J.}~\bibnamefont {Long}},
  \bibinfo {author} {\bibfnamefont {S.-T.}\ \bibnamefont {Wang}}, \bibinfo
  {author} {\bibfnamefont {N.}~\bibnamefont {Gemelke}},\ and\ \bibinfo {author}
  {\bibfnamefont {A.}~\bibnamefont {Keesling}},\ }\bibfield  {title} {\bibinfo
  {title} {Aquila: Quera's 256-qubit neutral-atom quantum computer},\ }\href
  {https://arxiv.org/abs/2306.11727} {\bibfield  {journal} {\bibinfo  {journal}
  {arXiv:2306.11727}\ } (\bibinfo {year} {2023})}\BibitemShut {NoStop}%
\bibitem [{\citenamefont {Pfeuty}(1970)}]{Pfeuty}%
  \BibitemOpen
  \bibfield  {author} {\bibinfo {author} {\bibfnamefont {P.}~\bibnamefont
  {Pfeuty}},\ }\bibfield  {title} {\bibinfo {title} {The one-dimensional ising
  model with a transverse field},\ }\href
  {https://doi.org/10.1016/0003-4916(70)90270-8} {\bibfield  {journal}
  {\bibinfo  {journal} {Ann. Phys.}\ }\textbf {\bibinfo {volume} {59}},\
  \bibinfo {pages} {79} (\bibinfo {year} {1970})}\BibitemShut {NoStop}%
\bibitem [{\citenamefont {Sachdev}(2011)}]{sachdev_2011}%
  \BibitemOpen
  \bibfield  {author} {\bibinfo {author} {\bibfnamefont {S.}~\bibnamefont
  {Sachdev}},\ }\href {https://doi.org/10.1017/CBO9780511973765} {\emph
  {\bibinfo {title} {Quantum Phase Transitions}}}\ (\bibinfo  {publisher}
  {Cambridge University Press},\ \bibinfo {address} {Cambridge, England},\
  \bibinfo {year} {2011})\BibitemShut {NoStop}%
\bibitem [{\citenamefont {Zamolodchikov}(1989{\natexlab{a}})}]{zam_1989}%
  \BibitemOpen
  \bibfield  {author} {\bibinfo {author} {\bibfnamefont {A.~B.}\ \bibnamefont
  {Zamolodchikov}},\ }\bibfield  {title} {\bibinfo {title} {Integrals of motion
  and s-matrix of the (scaled) $\textit{T} = \textit{T}_c$ ising model with
  magnetic field},\ }\href {https://doi.org/10.1142/S0217751X8900176X}
  {\bibfield  {journal} {\bibinfo  {journal} {Int. J. Mod. Phys. A}\ }\textbf
  {\bibinfo {volume} {04}},\ \bibinfo {pages} {4235} (\bibinfo {year}
  {1989}{\natexlab{a}})}\BibitemShut {NoStop}%
\bibitem [{\citenamefont {Fonseca}\ and\ \citenamefont
  {Zamolodchikov}(2003)}]{zam2003}%
  \BibitemOpen
  \bibfield  {author} {\bibinfo {author} {\bibfnamefont {P.}~\bibnamefont
  {Fonseca}}\ and\ \bibinfo {author} {\bibfnamefont {A.}~\bibnamefont
  {Zamolodchikov}},\ }\bibfield  {title} {\bibinfo {title} {Ising field theory
  in a magnetic field: Analytic properties of the free energy},\ }\href
  {https://link.springer.com/article/10.1023/A:1022147532606} {\bibfield
  {journal} {\bibinfo  {journal} {J. Stat. Phys}\ } (\bibinfo {year}
  {2003})}\BibitemShut {NoStop}%
\bibitem [{\citenamefont {Mussardo}(2010)}]{Mussardo:2010mgq}%
  \BibitemOpen
  \bibfield  {author} {\bibinfo {author} {\bibfnamefont {G.}~\bibnamefont
  {Mussardo}},\ }\href
  {https://doi.org/doi.org/10.1093/oso/9780198788102.001.0001} {\emph {\bibinfo
  {title} {{Statistical field theory}: {an introduction to exactly solved
  models in statistical physics}}}}\ (\bibinfo  {publisher} {Oxford Univ.
  Press},\ \bibinfo {address} {New York, NY},\ \bibinfo {year}
  {2010})\BibitemShut {NoStop}%
\bibitem [{\citenamefont {Eden}\ \emph {et~al.}(1966)\citenamefont {Eden},
  \citenamefont {Landshoff}, \citenamefont {Olive},\ and\ \citenamefont
  {Polkinghorne}}]{Eden:1966dnq}%
  \BibitemOpen
  \bibfield  {author} {\bibinfo {author} {\bibfnamefont {R.~J.}\ \bibnamefont
  {Eden}}, \bibinfo {author} {\bibfnamefont {P.~V.}\ \bibnamefont {Landshoff}},
  \bibinfo {author} {\bibfnamefont {D.~I.}\ \bibnamefont {Olive}},\ and\
  \bibinfo {author} {\bibfnamefont {J.~C.}\ \bibnamefont {Polkinghorne}},\
  }\href {https://inspirehep.net/literature/1517084} {\emph {\bibinfo {title}
  {{The analytic S-matrix}}}}\ (\bibinfo  {publisher} {Cambridge Univ. Press},\
  \bibinfo {address} {Cambridge},\ \bibinfo {year} {1966})\BibitemShut
  {NoStop}%
\bibitem [{\citenamefont {Zamolodchikov}\ and\ \citenamefont
  {Zamolodchikov}(1979)}]{ZAMOLODCHIKOV1979253}%
  \BibitemOpen
  \bibfield  {author} {\bibinfo {author} {\bibfnamefont {A.~B.}\ \bibnamefont
  {Zamolodchikov}}\ and\ \bibinfo {author} {\bibfnamefont {A.~B.}\ \bibnamefont
  {Zamolodchikov}},\ }\bibfield  {title} {\bibinfo {title} {Factorized
  {S}-matrices in two dimensions as the exact solutions of certain relativistic
  quantum field theory models},\ }\href
  {https://doi.org/https://doi.org/10.1016/0003-4916(79)90391-9} {\bibfield
  {journal} {\bibinfo  {journal} {Ann. Phys.}\ }\textbf {\bibinfo {volume}
  {120}},\ \bibinfo {pages} {253} (\bibinfo {year} {1979})}\BibitemShut
  {NoStop}%
\bibitem [{\citenamefont {Zamolodchikov}(1989{\natexlab{b}})}]{Zam_1989_IFT}%
  \BibitemOpen
  \bibfield  {author} {\bibinfo {author} {\bibfnamefont {A.~B.}\ \bibnamefont
  {Zamolodchikov}},\ }\bibfield  {title} {\bibinfo {title} {{Integrable field
  theory from conformal field theory}},\ }\href
  {https://doi.org/10.2969/aspm/01910641} {\bibfield  {journal} {\bibinfo
  {journal} {Adv. Stud. Pure Math.}\ }\textbf {\bibinfo {volume} {19}},\
  \bibinfo {pages} {641} (\bibinfo {year} {1989}{\natexlab{b}})}\BibitemShut
  {NoStop}%
\bibitem [{\citenamefont {Dorey}(1996)}]{Dorey_1996gd}%
  \BibitemOpen
  \bibfield  {author} {\bibinfo {author} {\bibfnamefont {P.}~\bibnamefont
  {Dorey}},\ }\bibfield  {title} {\bibinfo {title} {{Exact S matrices}},\ }in\
  \href@noop {} {\emph {\bibinfo {booktitle} {{Eotvos Summer School in Physics:
  Conformal Field Theories and Integrable Models}}}}\ (\bibinfo {year} {1996})\
  pp.\ \bibinfo {pages} {85--125},\ \Eprint
  {https://arxiv.org/abs/hep-th/9810026} {arXiv:hep-th/9810026} \BibitemShut
  {NoStop}%
\bibitem [{\citenamefont {Acerbi}\ \emph {et~al.}(1997)\citenamefont {Acerbi},
  \citenamefont {Mussardo},\ and\ \citenamefont {Valleriani}}]{CAcerbi_1997}%
  \BibitemOpen
  \bibfield  {author} {\bibinfo {author} {\bibfnamefont {C.}~\bibnamefont
  {Acerbi}}, \bibinfo {author} {\bibfnamefont {G.}~\bibnamefont {Mussardo}},\
  and\ \bibinfo {author} {\bibfnamefont {A.}~\bibnamefont {Valleriani}},\
  }\bibfield  {title} {\bibinfo {title} {On the form factors of relevant
  operators and their cluster property},\ }\href
  {https://doi.org/10.1088/0305-4470/30/9/007} {\bibfield  {journal} {\bibinfo
  {journal} {J. Phys. A: Math. Theor.}\ }\textbf {\bibinfo {volume} {30}},\
  \bibinfo {pages} {2895} (\bibinfo {year} {1997})}\BibitemShut {NoStop}%
\bibitem [{\citenamefont {Karowski}(1979)}]{KAROWSKI1979244}%
  \BibitemOpen
  \bibfield  {author} {\bibinfo {author} {\bibfnamefont {M.}~\bibnamefont
  {Karowski}},\ }\bibfield  {title} {\bibinfo {title} {On the bound state
  problem in 1+1 dimensional field theories},\ }\href
  {https://doi.org/https://doi.org/10.1016/0550-3213(79)90600-X} {\bibfield
  {journal} {\bibinfo  {journal} {Nucl. Phys. B}\ }\textbf {\bibinfo {volume}
  {153}},\ \bibinfo {pages} {244} (\bibinfo {year} {1979})}\BibitemShut
  {NoStop}%
\bibitem [{\citenamefont {Fring}\ \emph {et~al.}(1993)\citenamefont {Fring},
  \citenamefont {Mussardo},\ and\ \citenamefont {Simonetti}}]{city961}%
  \BibitemOpen
  \bibfield  {author} {\bibinfo {author} {\bibfnamefont {A.}~\bibnamefont
  {Fring}}, \bibinfo {author} {\bibfnamefont {G.}~\bibnamefont {Mussardo}},\
  and\ \bibinfo {author} {\bibfnamefont {P.}~\bibnamefont {Simonetti}},\
  }\bibfield  {title} {\bibinfo {title} {Form factors for integrable lagrangian
  field theories, the sinh-gordon model},\ }\href
  {https://doi.org/10.1016/0550-3213(93)90252-k} {\bibfield  {journal}
  {\bibinfo  {journal} {Nucl. Phys. B}\ }\textbf {\bibinfo {volume} {393}},\
  \bibinfo {pages} {413} (\bibinfo {year} {1993})}\BibitemShut {NoStop}%
\bibitem [{\citenamefont {Babujian}\ \emph {et~al.}(1999)\citenamefont
  {Babujian}, \citenamefont {Fring}, \citenamefont {Karowski},\ and\
  \citenamefont {Zapletal}}]{Babujian:1998uw}%
  \BibitemOpen
  \bibfield  {author} {\bibinfo {author} {\bibfnamefont {H.~M.}\ \bibnamefont
  {Babujian}}, \bibinfo {author} {\bibfnamefont {A.}~\bibnamefont {Fring}},
  \bibinfo {author} {\bibfnamefont {M.}~\bibnamefont {Karowski}},\ and\
  \bibinfo {author} {\bibfnamefont {A.}~\bibnamefont {Zapletal}},\ }\bibfield
  {title} {\bibinfo {title} {{Exact form-factors in integrable quantum field
  theories: The Sine-Gordon model}},\ }\href
  {https://doi.org/10.1016/S0550-3213(98)00737-8} {\bibfield  {journal}
  {\bibinfo  {journal} {Nucl. Phys. B}\ }\textbf {\bibinfo {volume} {538}},\
  \bibinfo {pages} {535} (\bibinfo {year} {1999})}\BibitemShut {NoStop}%
\bibitem [{\citenamefont {Babujian}\ and\ \citenamefont
  {Karowski}(2001)}]{Babujian:2001wp}%
  \BibitemOpen
  \bibfield  {author} {\bibinfo {author} {\bibfnamefont {H.}~\bibnamefont
  {Babujian}}\ and\ \bibinfo {author} {\bibfnamefont {M.}~\bibnamefont
  {Karowski}},\ }\bibfield  {title} {\bibinfo {title} {{The 'Bootstrap program'
  for integrable quantum field theories in (1+1)-dimension}},\ }in\ \href
  {https://doi.org/10.1142/9789812777478_0004} {\emph {\bibinfo {booktitle}
  {{From QCD to Integrable Models: Old Results and New Developments (In Honor
  of 70 Year Jubilee of Prof. S. Matinyan)}}}}\ (\bibinfo {year} {2001})\
  \Eprint {https://arxiv.org/abs/hep-th/0110261} {arXiv:hep-th/0110261}
  \BibitemShut {NoStop}%
\bibitem [{\citenamefont {Delfino}\ \emph {et~al.}(1996)\citenamefont
  {Delfino}, \citenamefont {Simonetti},\ and\ \citenamefont
  {Cardy}}]{DELFINO1996327}%
  \BibitemOpen
  \bibfield  {author} {\bibinfo {author} {\bibfnamefont {G.}~\bibnamefont
  {Delfino}}, \bibinfo {author} {\bibfnamefont {P.}~\bibnamefont {Simonetti}},\
  and\ \bibinfo {author} {\bibfnamefont {J.}~\bibnamefont {Cardy}},\ }\bibfield
   {title} {\bibinfo {title} {Asymptotic factorisation of form factors in
  two-dimensional quantum field theory},\ }\href
  {https://doi.org/https://doi.org/10.1016/0370-2693(96)01035-0} {\bibfield
  {journal} {\bibinfo  {journal} {Phys. Lett. B}\ }\textbf {\bibinfo {volume}
  {387}},\ \bibinfo {pages} {327} (\bibinfo {year} {1996})}\BibitemShut
  {NoStop}%
\bibitem [{\citenamefont {Delfino}\ and\ \citenamefont
  {Simonetti}(1996)}]{DELFINO1996450}%
  \BibitemOpen
  \bibfield  {author} {\bibinfo {author} {\bibfnamefont {G.}~\bibnamefont
  {Delfino}}\ and\ \bibinfo {author} {\bibfnamefont {P.}~\bibnamefont
  {Simonetti}},\ }\bibfield  {title} {\bibinfo {title} {Correlation functions
  in the two-dimensional ising model in a magnetic field at $\textit{T} =
  \textit{T}_c$},\ }\href
  {https://doi.org/https://doi.org/10.1016/0370-2693(96)00783-6} {\bibfield
  {journal} {\bibinfo  {journal} {Phys. Lett. B}\ }\textbf {\bibinfo {volume}
  {383}},\ \bibinfo {pages} {450} (\bibinfo {year} {1996})}\BibitemShut
  {NoStop}%
\end{thebibliography}%

\newpage
\onecolumngrid
\setcounter{figure}{0}
\makeatletter
\renewcommand{\thefigure}{S\@arabic\c@figure}
\setcounter{equation}{0} \makeatletter
\renewcommand \theequation{S\@arabic\c@equation}

\section{Supplemental Material---Berry connection and quantum geometry in time-dependent many-body systems with instantaneous quantum integrable field theory}

\section{A.~Introduction to integrable field theory}
\label{app:A}
In this section, we briefly summarize the necessary and basic knowledge of (1+1)-dimensional [(1+1)D] integrable field theory (IFT) to support the proof of our theorem.
We only focus on the IFT obtianed by a (1+1)D conformal field theory (CFT) perturbed by a relevant field $\mathcal{O}(x)$,
\begin{equation}
    \mathcal{H}_{\text{IFT}} = \mathcal{H}_{\text{CFT}} + \lambda \int \mathcal{O}(x) dx,
\end{equation}
with $\mathcal{H}_{\text{CFT}}$ and $\mathcal{H}_{\text{IFT}}$ denoting the Hamiltonian of a 1+1D CFT and its corresponding IFT, respectively~\cite{Zam_1989_IFT,zam_1989, fateev, Mussardo:2010mgq}. 
This Hamiltonian possesses an infinite number of local conserved quantities, 
where the corresponding conserved charges $\mathscr{Q}_s$,  obtained by integrating the local conserved quantities, satisfy
\begin{equation}
    \left[\mathcal{H}_{\text{IFT}}, \mathscr{Q}_s\right] = 0,~~\left[\mathscr{Q}_s, \mathscr{Q}_{s'}\right] = 0,
\end{equation}
where $s$ is the Lorentz spin of the conserved charge, borrowed from the language of the original CFT.
The relevant perturbation generates massive excitations of the IFT.
Furthermore, such massive excitations are generally described by multi-quasiparticle spectra with relativistic dispersion $E_{A_i}^2 = p_{A_i}^2 + m_{A_i}^2$. 
The latter can be re-parametrized by the rapidity $\vartheta$ as
$E_{A_i} = m_{A_i}\cosh\vartheta_i,~p_{A_i} = m_{A_i}\sinh \vartheta_i$.
Here $A_i$ is the type species of the particle~\cite{KAROWSKI1979244,ZAMOLODCHIKOV1979253,Dorey_1996gd, Eden:1966dnq}. 
The asymptotic $N$-particle states can then be written as
\begin{equation}
    \wec{\psi_{N}}=\wec{A_{1}(\vartheta_1)...A_{N}(\vartheta_N)},
    \label{Eq:supp_istate}
\end{equation}
with the normalization condition $\langle A_{i}(\vartheta_i)|A_{j}(\vartheta_j)\rangle=2\pi\delta_{A_{i}A_{j}}\delta(\vartheta_i-\vartheta_{j})$. 
The matrix element of local fields $\mathcal{O}$ between the vacuum (ground state) and the multi-particle states $\wec{\psi_{N}}$, known as the elementary form factor (FF), is defined as 
\begin{equation}
    F^{\mathcal{O}}_{A_1,\ldots,A_N}(\vartheta_1,\ldots,\vartheta_N)
    \equiv \mathcal{F}^{\mathcal{O}}_{N}(\vartheta_1,\ldots,\vartheta_N)
    = \langle 0|\mathcal{O}(0)\wec{\psi_{N}} = \langle 0|\mathcal{O}(0)\wec{A_{1}(\vartheta_1)...A_{N}(\vartheta_N)},
\end{equation}
where we have abbreviated the particle types as the total particle number $N$.
The FF can be systematically computed through FF bootstrap method~\cite{delfino_1995, DELFINO1996450,H_ds_gi_2019, smirnov_1992, Babujian:1998uw, city961, Babujian:2001wp, CAcerbi_1997,DELFINO1996327}.
A general matrix element (FF) of $\mathcal{O}$ can be obtained from elementary FFs using the crossing relation~\cite{delfino_1995,H_ds_gi_2019,xiao_2021,xiao_2024time}
\begin{equation}
\begin{aligned}
\mathcal{F}_{NM}^{\mathcal{O}}(\vartheta'_{1},...,\vartheta'_{N}\vert\vartheta_{1},...,\vartheta_{M})&=\cew{A_{1}(\vartheta'_1)...A_{N}(\vartheta'_N)}\mathcal{O}\wec{A_{1}(\vartheta_1)...A_{M}(\vartheta_M)}
\\&=\mathcal{F}_{N-1;M+1}^{\mathcal{O}}(\vartheta'_{1},...,\vartheta'_{N-1}\vert\vartheta'_{N}+i\pi,\vartheta_{1},...,\vartheta_{M}) +\sum_{k=1}^{M}
\Bigg[ 2\pi\delta_{A_{N}A_{k}}\delta(\vartheta'_{N}-\vartheta_{k})
\\
&\times \prod_{l=1}^{k-1}S_{A_{l}A_{k}}(\vartheta_{l}-\vartheta_{k}) \mathcal{F}^{\mathcal{O}}_{N-1;M-1}(\vartheta'_{1},...,\vartheta'_{N-1}\vert\vartheta_{1},...,\vartheta_{k-1},\vartheta_{k+1},...,\vartheta_{M})
\Bigg],
\label{Eq:cross}
\end{aligned}
\end{equation} 
with $S_{A_{l}A_{k}}(\vartheta_{l}-\vartheta_{k})$ being the corresponding scattering matrix.
Such IFT can further be considered on a finite system with sufficiently large spatial length $L$, while time ranges from $-\infty$ to $+\infty$, known as the finite-size formalism for IFT. 
This formalism was originally developed by B. Pozsgay and G. Tak{\'a}cs, with great details in~\cite{POZSGAY2008167,POZSGAY2008209,H_ds_gi_2019}.
Within the finite-size framework, the $N$-particle asymptotic states follow
\begin{equation}
    \wec{\psi_{N}}_{L}=\wec{A_{1}(I_1)...A_{N}(I_N)}_{L},
    \label{Eq:supp_Lstate}
\end{equation}
with discrete quantum number $I_i$ and usual normalization condition $\langle A_{i}(I_i)|A_{j}(I_j)\rangle=\delta_{A_{i}A_{j}}\delta_{I_{i}I_{j}}$.
Considering the periodic boundary conditions, these quantum numbers are connected with the rapidities via the Bethe-Yang equation,
\begin{equation}
    m_{A_{i}}L\sinh\vartheta_{i}-i\sum_{A_{j}\neq A_{i}}\ln S_{A_{i}A_{j}}(\vartheta_{i}-\vartheta_{j}) = 2\pi I_{i} = Q_{A_{i}},
    \label{Eq:supp_BYeq}
\end{equation}
Furthermore, it is clearly observed that 
there is a measure difference between the normalizations of the two quantum states in different framework. 
All the related quantities such as the complete basis and form factors (FF) will then be influenced. 
This issue can be done by introducing the Jacobian given by the density of states of the corresponding particle sets,
\begin{equation}
    \rho_{L,\{\mathscr{A}\}_{N}}(\{\vartheta\}_{N})=\rho_{L,\{\mathscr{A}\}_{N}}= \text{Det}  \left( \frac{\partial Q_{A_k}(\{\vartheta\}_{N})}{\partial \vartheta_{l}} \right),
    \label{Eq:supp_jacobian}
\end{equation}
with $\{\mathscr{A}\}_{N}=\{A_{1},...,A_{N}\}$ and $\{\vartheta\}_{N}=\{\vartheta_1,\ldots,\vartheta_N\}$ 
being the abbreviation of particle sets and the set of rapidities of correspond particles. The abbreviations
are also applied in the following sections.
We then have the following measure transformation between all the associated quantities,
\begin{equation}
\begin{aligned}
\langle 0 &\vert \mathcal{O} \vert A_{1}(I_{1})...A_{N}(I_{N})\rangle_{L} = \frac{\mathcal{F}^{\mathcal{O}}_{N}(\{\vartheta\}_{N})}{\sqrt{\rho_{L,\{\mathscr{A}\}_{N}} } }, \\
&\vert A_{1}(I_{1})...A_{N}(I_{N})\rangle_{L} = \frac{\vert A_{1}(\vartheta_{1})...A_{N}(\vartheta_{N})\rangle}{\sqrt{\rho_{L,\{\mathscr{A}\}_{N}} } },\\
&\sum_{I_{1},...,I_{N}}=\int\frac{d\vartheta_{1}...d\vartheta_{N}}{(2\pi)^N}\rho_{L,\{\mathscr{A}\}_{N}}.
\label{Eq:supp_measurechange}
\end{aligned}
\end{equation}

\section{B.~Proof of the theorem}
\label{app:B}
This section gives the proof of the theorem in the main text. Considering an adiabatic transformation of an eigenstate in 
an IFT containing $n$ stable quasiparticles, 
\begin{equation}
    \wec{\psi_{N}(t)}=\mathscr{U}(t)\wec{\psi_{N}}_{\text{init},L},
    \label{Eq:supp_quench}
\end{equation}
where we have put this quantum field theory on a cylinder with circumference $L$, with the adiabatic transformation satisfying
\begin{equation}
i\mathscr{U}^{-1}(t)\dot{\mathscr{U}}(t)=\sum_{i=1}^{N_{\mathcal{O}}}\int_{0}^{L}f_{i}(t)\mathcal{O}_{i}(x)dx,
\label{Eq:supp_W}
\end{equation} 
which is also known as the effective adiabatic Hamiltonian~\cite{yongdezhang}. The initial state is given by
\begin{equation}
\wec{\psi_{N}}_{\text{init},L}=\wec{A_{1}(\vartheta_1;I_1)...A_{N}(\vartheta_N;I_N)}_{L}.    
\end{equation}
The related parameters have been introduced in the main text. According to the definition of the Berry connections, we focus on calculating the quantities $\gamma_{NM} = i \langle\psi_N(t)\vert\dot{\psi}_M(t)\rangle$. 
Introducing $\vert\psi_M(t)\rangle$ with $M$ stable quasi-particles, then we have~\cite{H_ds_gi_2019}
\begin{equation}
    \gamma_{NM}=L\sum_{i=1}^{N_{\mathcal{O}}}f_{i}(t)\langle\psi_N\vert\mathcal{O}_{i}\vert\psi_M\rangle|_{p_{\psi_N}=p_{\psi_M}},
\end{equation}
where $p_{\psi_N}$ and $p_{\psi_M}$ are the total momentum of the corresponding quantum states. Using the framework outlined in Sec.~A, let us consider a specific $\mathcal{O}_{i}$: Applying crossing relation and the formula given in Eq.~(\ref{Eq:supp_measurechange}), we obtain the following expression for the off-diagonal matrix element
\begin{equation}
\begin{aligned}
\langle\psi_N\vert\mathcal{O}_{i}\vert\psi_M\rangle={}_{L}\cew{A_{1}(\vartheta'_1;I'_1)...A_{N}(\vartheta'_N;I'_N)}\mathcal{O}_{i}\wec{A_{1}(\vartheta_1;I_1)...A_{M}(\vartheta_M;I_M)}_{L}
=\frac{\mathcal{F}_{NM}^{\mathcal{O}_i}(\{\vartheta'+i\pi\}_{N},\{\vartheta\}_{M})}{\sqrt{\rho_{L,\{\mathscr{A'}\}_{N}}}\sqrt{\rho_{L,\{\mathscr{A}\}_{M}}}}.
\label{Eq:supp_nd}    
\end{aligned}
\end{equation}
Eq.~(\ref{Eq:cross}) shows that when one
particle in the bra carries on the same rapidity as one particle in the ket, 
there will be a divergence encoded in the delta function. 
The strongest divergence happens for the diagonal matrix elements
which we shall pay more attention to carefully 
deal with~\cite{POZSGAY2008167,POZSGAY2008209}. A way to correctly recover such divergence in the finite size 
framework of IFT is developed in~\cite{POZSGAY2008167,POZSGAY2008209,H_ds_gi_2019},
which follows,
\begin{equation}
{}_{L}\cew{A_{1}(\vartheta_1;I_1)...A_{N}(\vartheta_N;I_N)}\mathcal{O}_{i}\wec{A_{1}(\vartheta_1;I_1)...A_{N}(\vartheta_N;I_N)}_{L}\approx{}_{L}\langle\mathcal{O}_{i}\rangle_{L}+\sum_{\forall \mathscr{S}\subseteq \{\mathscr{A}\}_{N}}\frac{\mathcal{F}_{\mathscr{S}\mathscr{S}}^{\mathcal{O}_{i}}(\{\vartheta+i\pi\}_{\mathscr{S}},\{\vartheta\}_{\mathscr{S}})\rho_{L,\{\mathscr{A}\}_{N}\setminus \mathscr{S}}}{\rho_{L,\{\mathscr{A}\}_{N}}},
\label{Eq:supp_finiteFF}
\end{equation}
where $\mathscr{S}$ is a non-empty subset of $\{\mathscr{A}\}_{N}$, 
and the summation runs over all non-empty particle subsets. 
$\mathcal{F}_{\mathscr{S}\mathscr{S}}^{\mathcal{O}_{i}}(\{\vartheta+i\pi\}_{\mathscr{S}},\{\vartheta\})_{\mathscr{S}}$ 
is the FF where both the bra and ket are of particle set $\mathscr{S}$ with corresponding rapidity set $\{\vartheta+i\pi\}_{\mathscr{S}}$ and $\{\vartheta\}_{\mathscr{S}}$. 
$\{\mathscr{A}\}_{N}\setminus \mathscr{S}$ is the particle set that excludes $\mathscr{S}$ from $\{\mathscr{A}\}_{N}$.
By making use of Eq.~(\ref{Eq:supp_nd}) and Eq.~(\ref{Eq:supp_finiteFF}), we obtain the results for the Berry connection matrix entries. Finally, for the off-diagonal matrix entries that $M\neq N$, or $M= N$ but not all the rapidities in the bra and ket are the same, 
\begin{equation}
    \frac{\gamma_{NM}}{L}\approx \sum_{i}f_i(t)\frac{\mathcal{F}_{NM}^{\mathcal{O}_i}(\{\vartheta'+i\pi\}_{N},\{\vartheta\}_{M})}{\sqrt{\rho_{L,\{\mathscr{A'}\}_{N}}}\sqrt{\rho_{L,\{\mathscr{A}\}_{M}}}}.
    \label{Eq:supp_gammand}
\end{equation}
For the diagonal matrix elements that $M=N$ and all the rapidities are the same in the bra and ket, 
\begin{equation}
\begin{aligned}
    \frac{\gamma_{NN}}{L}\approx\sum_{i}f_{i}(t)\bigg{[}{}_{L}\langle\mathcal{O}_{i}\rangle_{L}+
    \sum_{\forall \mathscr{S}\subseteq \{\mathscr{A}\}_{N}}\frac{\mathcal{F}_{\mathscr{S}\mathscr{S}}^{\mathcal{O}_{i}}(\{\vartheta+i\pi\}_{\mathscr{S}},\{\vartheta\}_{\mathscr{S}})\rho_{L,\{\mathscr{A}\}_{N}\setminus \mathscr{S}}}{\rho_{L,\{\mathscr{A}\}_{N}}}\bigg{]},
\end{aligned}
\label{Eq:supp_gammad}
\end{equation}
as shown in the main text. Noticing that the Jacobian Eq.~(\ref{Eq:supp_jacobian}) is a polynomial of order $L^N$,
thus providing the asymptotic algebraic behaviour of $\gamma_{NM}(L)$ being $L^{1-\mathcal{Z}/2}$ with $\mathcal{Z}=M+N$ being the total number of involved particles and $L\rightarrow\infty$.
Following this conclusion, we can observe that in Eq.~(\ref{Eq:supp_gammand}) only the term that $(M+N)/2 \leq 1$ [as in Eq.~(\ref{Eq:supp_gammad}) 
only the term that $\#$  $\leq 1$] finally survives in the thermodynamic limit ($L\rightarrow\infty$), 
suggesting that the contributions for the Berry connection matrix elements are up to two particle excitations.

\section{C.~The instantaneous energy levels and eigenstates}
\label{app:C}
We investigate the instantaneous \schro~equation $\mathcal{H}(t)\wec{\psi_{n}(t)}_{ins}=E_{n}(t)\wec{\psi_{n}(t)}_{ins}$ for the following Hamiltonian,
\begin{equation}
    \mathcal{H}(t)=-J\sum_{i=1}^{N}\left(\sigma^{z}_{i}\sigma^{z}_{i+1}+\cos\w_0 t\sigma^{x}_{i}-\sin\w_0 t\sigma^{y}_{i}+h_{z}\sigma^{z}_{i}\right).
    \label{Eq:supp_latH_driven}
\end{equation}
Under the unitary transformation $\mathcal{U}(t)=\exp\left\lbrace -\frac{i}{2}\w_{0} t\sum_{i}\sigma_{i}^{z}\right\rbrace$, we have~\cite{Robinson,xiao_2024time}
\begin{equation}
    \mathcal{U}(t)\mathcal{H}(t)\mathcal{U}(t)^{-1}=-J\sum_{i=1}^{L}\left(\sigma^{z}_{i}\sigma^{z}_{i+1}+\sigma^{x}_{i}+h_{z}\sigma^{z}_{i}\right) = \mathcal{H}_{t=0}.
    \label{Eq:supp_Utrans}
\end{equation}
Applying the eigenstate of $\mathcal{H}_{t=0}$ on both side,
\begin{equation}
\mathcal{U}(t)\mathcal{H}(t)\mathcal{U}(t)^{-1}\wec{\psi_{n}(0)}=\mathcal{H}_{t=0}\wec{\psi_{n}(0)}=E_n\wec{\psi_{n}(0)},
\end{equation}
and this further gives
\begin{equation}
\mathcal{H}(t)\mathcal{U}(t)^{-1}\wec{\psi_{n}(0)}=E_n\mathcal{U}(t)^{-1}\wec{\psi_{n}(0)},
\end{equation}
i.e., the instantaneous eigenstate $\wec{\psi_{n}(t)}_{ins}$ is proportional to $\mathcal{U}(t)^{-1}\wec{\psi_{n}(0)}$ while they have the same time-independent eigenenergy $E_n$.
Without loss of generality, we conclude
\begin{equation}
    E_n(t)=E_n(0),~~\wec{\psi_{n}(t)}_{ins}=e^{i\varphi_{n}(t)}\mathcal{U}(t)^{-1}\wec{\psi_{n}(0)},
    \label{Eq:supp_inswavefunction}
\end{equation}
with $\varphi_{n}(t)$ being a time-dependent U(1) gauge.

\section{D.~Calculation of the quantum geometric potential}
\label{app:D}
The effective energy gap is defined as $\Delta_{mn}(t)=E_m(t)-E_n(t)+Q_{nm}(t)$,
where $E_{m,n}(t)$ are the instantaneous eigenenergy of the corresponding states, and
\begin{equation}
    Q_{nm}(t)=\gamma_{nn}(t)-\gamma_{mm}(t)+\frac{d}{dt}\text{arg}\left[\langle\psi_{m}(t)\vert\dot{\psi_{n}}(t)\rangle\right]_{ins}
    \label{Eq:supp_QGP}
\end{equation}
is known as the quantum geometric potential (QGP), with $\gamma_{nn}(t)={}_{ins}\langle\psi_{n}(t)\vert i\partial_{t}\vert\psi_{n}(t)\rangle_{ins}$ 
being the Berry connection~\cite{da_QGP2008,da_QGP2018,da_tunneling}. We consider the effective energy gap between any two arbitrary excited states, which in the scaling limit corresponds to two multi-$E_8$ particle states containing $N$ and $M$ quantum $E_8$ particles, respectively. The instantaneous energy gap is proven to be static, and can be directly read as
\begin{equation}
    E_{m}(t)-E_{n}(t)=\sum_{\beta=1}^{M}m_{\beta}\cosh\vartheta_{\beta}-\sum_{\alpha=1}^{N}m_{\alpha}\cosh\vartheta'_{\alpha}.
    \label{Eq:supp_insEnergygap}
\end{equation}
On the lattice, using Eq.~(\ref{Eq:supp_inswavefunction}), the Berry connection is given by
\begin{equation}
\begin{aligned}
    \gamma_{mm}(t)&=-\dot{\varphi}_{m}(t)-\frac{\w_0}{2}\sum_{i=1}^{N}\cew{\psi_{m}(0)}\sigma^{z}_{i}\wec{\psi_{m}(0)},\\
    \gamma_{nn}(t)&=-\dot{\varphi}_{n}(t)-\frac{\w_0}{2}\sum_{i=1}^{N}\cew{\psi_{n}(0)}\sigma^{z}_{i}\wec{\psi_{n}(0)}.
    \label{EqLsupp_BerryConnection}
\end{aligned}
\end{equation}
Since we also have
\begin{equation}
   \frac{d}{dt}\text{arg}\left[\langle\psi_{m}(t)\vert\dot{\psi_{n}}(t)\rangle\right]_{ins}= \dot{\varphi}_{n}(t)-\dot{\varphi}_{m}(t).
   \label{Eq:supp_phasearg}
\end{equation}
The QGP is given as following,
\begin{equation}
    Q_{nm}(t)=\frac{\w_0}{2}\sum_{i=1}^{N}\left(\cew{\psi_{m}(0)}\sigma^{z}_{i}\wec{\psi_{m}(0)}-\cew{\psi_{n}(0)}\sigma^{z}_{i}\wec{\psi_{n}(0)}\right),
    \label{Eq:supp_QGPlattice}
\end{equation}
which also implies that the QGP is U(1) gauge invariant. To obtain the exact expression of the QGP, we use the finite size field theory frame and the results in Sec.~A and B~\cite{POZSGAY2008167,POZSGAY2008209,H_ds_gi_2019}. We first transform the system into the scaling limit. Since $\langle\sigma\rangle=\bar{s}J^{1/8}\langle\sigma^{z}\rangle$~\cite{zam2003,xiao_2024time},
\begin{equation}
    Q^{s}_{nm}=\frac{\w_0}{2}\bar{s}^{-1}J^{-1/8}\lim_{L\rightarrow\infty}2JL\left({}_{L}\cew{\vartheta_{1}...\vartheta_{M}}\sigma\wec{\vartheta_{1}...\vartheta_{M}}_{L}-{}_{L}\cew{\vartheta'_{1}...\vartheta'_{N}}\sigma\wec{\vartheta'_{1}...\vartheta'_{N}}_{L}\right),
    \label{Eq:supp_QGPfield}
\end{equation}
with $L=Na$ being the length of the system, and $a$ being the lattice spacing, satisfying $2Ja=\hbar c=1$~\cite{delfino_1995,xiao_2021}. Introducing the general FF for the quantum $E_8$ field theory,
\begin{equation}
    F^{\mathcal{O}}_{a_{1},...,a_{n},b_{1},...,b_{m}}(\vartheta_{1},...,\vartheta_{n},\vartheta_{n+1},...,\vartheta_{n+m})
={}_{a_{1},...,a_{n}}\langle \vartheta_{1},...,\vartheta_{n}\vert \mathcal{O}\vert \vartheta_{n+1},...,\vartheta_{n+m}\rangle_{b_{1},...,b_{m}}/\langle\mathcal{O}\rangle.
\label{Eq:supp_FF}
\end{equation}
The $E_{8}$ FF theory and a complete FF bootstrapping process can be found in Refs.~\cite{delfino_1995,H_ds_gi_2019,xiao_2021}. 
According to previous analysis of the FFs, the QGP follows,
\begin{equation}
    Q^{s}_{nm}(t)=\frac{\w_0}{2}\bar{s}^{-1}J^{-1/8}\langle\sigma\rangle 2J\left(\sum_{\beta=1}^{M}\frac{F^{\sigma}_{\beta\beta}(i\pi,0)}{m_{\beta}\cosh\vartheta_{\beta}}-\sum_{\alpha=1}^{N}\frac{F^{\sigma}_{\alpha\alpha}(i\pi,0)}{m_{\alpha}\cosh\vartheta'_{\alpha}}\right).
    \label{Eq:supp_QGP2}
\end{equation}
This result implies that the QGP for this system is only related with the diagonal two-particle FFs in quantum $E_8$ field theory. 
In fact, it can be further reduced to a more transparent form. Since $m_{\beta}=C^{s}_{m}h^{8/15}e_{\beta}$, 
with $e_{\beta}$ being the Cartan roots of $E_8$ exceptional Lie algebra, $\langle\sigma\rangle=C^{s}_{\sigma}h^{1/15}$, 
with $h$ being $h_{z}$ in the scaling limit, and $C^{s}_{\sigma}=4(C^{s}_{m})^2/15\varphi_{11}$, where $\varphi_{11}$ is a constant~\cite{delfino_1995}, we can obtain that,
\begin{equation}
    \begin{aligned}
        Q^{s}_{nm}(t)&=\frac{\w_0}{2}\bar{s}^{-1}J^{-1/8}\langle\sigma\rangle 2J\left(\sum_{\beta=1}^{M}\frac{F^{\sigma}_{\beta\beta}(i\pi,0)}{m_{\beta}\cosh\vartheta_{\beta}}-\sum_{\alpha=1}^{N}\frac{F^{\sigma}_{\alpha\alpha}(i\pi,0)}{m_{\alpha}\cosh\vartheta'_{\alpha}}\right) \\
        &=\frac{\w_0}{2}\frac{4C^{s}_{m}}{15\varphi_{11}}\bar{s}^{-1}J^{-1/8} 2Jh^{-7/15}\left(\sum_{\beta=1}^{M}\frac{F^{\sigma}_{\beta\beta}(i\pi,0)}{m_{\beta}\cosh\vartheta_{\beta}}-\sum_{\alpha=1}^{N}\frac{F^{\sigma}_{\alpha\alpha}(i\pi,0)}{m_{\alpha}\cosh\vartheta'_{\alpha}}\right). 
    \label{Eq:supp_QGP3}
    \end{aligned}
\end{equation}
It can be proven that $F^{\sigma}_{\beta\beta}(i\pi,0)/e^{2}_{\beta}=-2\varphi_{11}$, 
which is not related with specific particle type. For showing this, 
we consider the trace of the stress-energy tensor $\Theta(x)$~\cite{delfino_1995},
\begin{equation}
    \cew{m_{\beta}}\Theta(x)\wec{m_{\beta}}=2\pi m_{\beta}^{2}=2\pi (C^{s}_{m})^{2} h^{16/15} e_{\beta}^2.
    \label{Eq:supp_TTtrace}
\end{equation}
On the other hand, since $\Theta(x)=2\pi h(2-2\Delta_{\sigma})\sigma(x)$, where $\Delta_{\sigma}=1/16$ is the conformal weight of $\sigma(x)$, together with $C^{s}_{\sigma}=4(C^{s}_{m})^2/15\varphi_{11}$, we can immediately obtain that $F^{\sigma}_{\beta\beta}(i\pi,0)/e^{2}_{\beta}=-2\varphi_{11}$. Then the QGP is further reduced as
\begin{equation}
    \begin{aligned}
        Q^{s}_{nm}(t)&=-\frac{\w_0}{2}\frac{8}{15h}\bar{s}^{-1}J^{-1/8} 2J\left(\sum_{\beta=1}^{M}\frac{m_{\beta}}{\cosh\vartheta_{\beta}}-\sum_{\alpha=1}^{N}\frac{m_{\alpha}}{\cosh\vartheta'_{\alpha}}\right).
        \label{Eq:supp_QGPscalinglimit}
    \end{aligned}
\end{equation}
Sending back to the lattice limit, using $h=2J^{15/8}h_{z}/\bar{s}$~\cite{xiao_2021,xiao_2024time}, we finally obtain that
\begin{equation}
    Q_{nm}(t)=-\frac{\w_0}{2J}\frac{8}{15h_{z}}\left(\sum_{\beta=1}^{M}\frac{m_{\beta}}{\cosh\vartheta_{\beta}}-\sum_{\alpha=1}^{N}\frac{m_{\alpha}}{\cosh\vartheta'_{\alpha}}\right).
    \label{Eq:supp_QGPlattice}
\end{equation}
This implies that the QGPs in this system are only related with the contributions from all the quantum $E_8$ particles 
in the corresponding quantum $E_8$ states.

\section{E.~Technical Details of TLFFA}
\label{app:E}
We provide additional details about the truncated lattice free fermion approach (TLFFA). Given the Hamiltonian
\begin{equation}
    \mathcal{H} =  \mathcal{H}_0 - J h_z \sum_{i = 1}^{L} \sigma_{i}^{z},
    \label{Eq:supp_H_E8}
\end{equation}
where $ \mathcal{H}_{0}$ represents the Hamiltonian of the transverse field Ising chain with periodic boundary condition,
\begin{equation}
    \mathcal{H}_0 = -J\sum_{i=1}^{L}\left(\sigma^{z}_{i}\sigma^{z}_{i+1}+g\sigma^{x}_{i}\right).
    \label{Eq:supp_H_Ising}
\end{equation}
    
\noindent To construct the Hamiltonian for TLFFA, we introduce the dual Jordan-Wigner (JW) transformation ~\cite{Iorgov2011}. The usual JW transformation is given by~\cite{Pfeuty,sachdev_2011}
\begin{equation}
    \sigma_{i}^{z} = \prod_{i'<i} \left(1 - 2 c_{i'}^{\dagger}c_{i'}\right) \left(c_i^{\dagger} + c_i\right), ~~\sigma_i^{x} = 1-2c_i^{\dagger}c_i.
    \label{Eq:supp_JW}
\end{equation}
Thus the Hamiltonian will be transformed into
\begin{equation}
    \mathcal{H}_0 = -J \sum_{i = 1}^{L} \left[ \left( c_i^{\dagger} - c_i \right) \left( c_{i+1}^{\dagger} + c_{i+1} \right) - g \left( c_i^{\dagger} - c_i \right) \left( c_i^{\dagger} + c_i \right) \right].
    \label{Eq:supp_H_JW}
\end{equation}
The dual JW transformation is then given by~\cite{Iorgov2011}
\begin{equation}
    c_{i}^{\dagger} - c_{i} = -(a_{i}^{\dagger} - a_{i}), ~~ c_{i+1}^{\dagger} + c_{i+1} = a_{i}^{\dagger} + a_{i},
    \label{Eq:supp_JW_dJW}
\end{equation}
with $a_j^{\dagger}$ and $a_j$ being dual JW fermion creation and annihilation
operators, and $ \mathcal{H}_{0}$ follows
\begin{equation}
    \mathcal{H}_0 = 2J \sum_{i=1}^{L} \left(a_i^{\dagger} a_i - \frac{1}{2}\right) -Jg \sum_{i=1}^{L} \left(a_{i+1}^{\dagger} -a_{i+1}\right)\left(a_{i}^{\dagger} +a_{i}\right).
    \label{Eq:supp_H_Ising_fermion}
\end{equation}
The corresponding Hilbert space is expressed as a direct sum of the Ramond (R, $\mathcal{V}_{R}$) and Neveu-Schwarz (NS, $\mathcal{V}_{NS}$) sectors. We introduce the Bogoliubov transformation~\cite{sachdev_2011}
\begin{equation}
    \gamma_q = u_q a_q - i v_q a_{-q}^{\dagger},~~\gamma_q^{\dagger} = u_q a_q^{\dagger} + i v_q a_{-q},
\end{equation}
with
\begin{equation}
    u_q = \cos \frac{\theta_q}{2},~~v_q = \sin \frac{\theta_q}{2},~~ \tan \theta_q= \frac{g \sin q}{1- g \cos q},
\end{equation}
in particular for $q = 0$,
\begin{equation}
    u_q = 1,~~ v_q = 0.
\end{equation}
$ \mathcal{H}_{0}$ can then be diagonalised as
\begin{equation}
    \mathcal{H}_0 = \sum_{q} \varepsilon_q \left( \gamma_q^{\dagger}\gamma_q - \frac{1}{2}\right),
\end{equation}
\noindent where
\begin{equation}
    \varepsilon_q = 2J \sqrt{1+g^2 - 2g\cos q}.
    \label{Eq:supp_eplison}
\end{equation}
The eigenvectors in the two subspaces are
\begin{align}
    &\wec{P}_{R} = \wec{p_1,\dots,p_n}_{R} = \gamma_{p_1}^{\dagger}\gamma_{p_2}^{\dagger}\gamma_{p_3}^{\dagger} \dots \gamma_{p_n}^{\dagger} \wec{0}_{R} \in \mathcal{V}_{R},\\
    &\wec{Q}_{NS} = \wec{q_1,\dots,q_m}_{NS} = \gamma_{q_1}^{\dagger}\gamma_{q_2}^{\dagger}\gamma_{q_3}^{\dagger} \dots \gamma_{q_m}^{\dagger} \wec{0}_{NS} \in \mathcal{V}_{NS},
    \label{Eq:supp_NS_basis}
\end{align}
with the quantum number $p$ and $q$ 
\begin{equation}
    \mathcal{P}: \left\{p = \frac{2\pi }{L}j \right\},~~\mathcal{Q}: \left\{q = \frac{2\pi }{L}\left(j-\frac{1}{2}\right)\right\},~~j=-\frac{L}{2}+1,-\frac{L}{2}+2,\dots,\frac{L}{2}.
\end{equation}
The non-trivial part for the construction of Eq.~(\ref{Eq:supp_H_E8}) is given by the FF of $\sigma^{z}$ operator, and the FF of $\sigma^{z}_{1}$ follows~\cite{Iorgov2011}
\begin{equation}
    \begin{aligned}
        {}_{R}{\cew{P}} \sigma_1^{z}\wec{Q}_{NS} &= {}_{R}{\cew{p_1,p_2,\dots,p_n}} \sigma_1^{z}\wec{q_1,q_2,\dots,q_m}_{NS} \\
        &= \delta_{m-n,0}^{(\text{mod}\ 2)} i^{-(n+m)/2} (-1)^{n(n-1)/2} \prod_{p \in P} e^{-ip/2} \prod_{q \in Q} e^{iq/2}\left( \frac{2}{L} \right)^{(m+n)/2} h^{(m-n)^2 /4} \sqrt{\xi \xi_T}  \\ 
        &\quad \times\prod_{q \in Q} \frac{e^{\eta(q)/2}}{\sqrt{\varepsilon_q}} \prod_{p \in P} \frac{e^{-\eta(p)/2}}{\sqrt{\varepsilon_p}} \prod_{q < q' \in Q} \frac{4 \sin \left( \frac{q - q'}{2} \right)}{\varepsilon_q + \varepsilon_{q'}} \prod_{p < p' \in P} \frac{4 \sin \left( \frac{p - p'}{2} \right)}{\varepsilon_p + \varepsilon_{p'}} \prod_{q \in Q} \prod_{p \in P} \frac{\varepsilon_q + \varepsilon_p}{4 \sin \left( \frac{q - p}{2} \right)},
    \end{aligned}
    \label{Eq:supp_FF_Sx_ladder_1}
\end{equation}
where $\xi = (1-g^2)^{1/4}$ and
\begin{equation}
    \begin{aligned}
        \xi_T = \frac{\prod_{q \in \mathcal{Q}} \prod_{p \in \mathcal{P}} (\varepsilon_q + \varepsilon_p)^{1/2}}{\prod_{q,q' \in \mathcal{Q}} (\varepsilon_q + \varepsilon_{q'})^{1/4} \prod_{p,p' \in \mathcal{P}} (\varepsilon_p + \varepsilon_{p'})^{1/4}},~~e^{\eta(\alpha)} = \frac{\prod_{q' \in \mathcal{Q}} (\varepsilon_\alpha + \varepsilon_{q'})}{\prod_{p' \in \mathcal{P}} (\varepsilon_\alpha + \varepsilon_{p'})}.
    \end{aligned}
    \label{Eq:supp_xiT_eEta}
\end{equation}
The spin operator $\sigma_l^{z}$ is then associated with $\sigma_{1}^{z}$ as
\begin{equation}
	{}_{R}{\cew{P}} \sigma_l^{z}\wec{Q}_{NS} 
	= e^{i (l-1) \left(\sum_{q\in Q} q - \sum_{p \in P} p\right)} {}_{R}{\cew{P}} \sigma_1^{z}\wec{Q}_{NS}.
    \label{Eq:supp_FF_Sx_ladder_j}
\end{equation}
We can then obtain all the matrix elements for Eq.~(\ref{Eq:supp_H_E8}). However, due to the exponential growth of the Hilbert space, a truncation of the full Hamiltonian is necessary, which is the key to TLFFA.
In this paper, we include only states that satisfy the zero total momentum constraint and have an energy upper bound of $10J$~\cite{zam_1990,Tu2410},
\begin{equation}
    \begin{aligned}
        \sum_{i} k_i = 0\; {\text{and}} \;  \sum_{k_i} \varepsilon_{k_i} \le 10J.
    \end{aligned}
    \label{Eq:supp_truncated_condition}
\end{equation}

\section{F.~LE spectrum}
\label{app:F}
The detailed data for the LE spectral entropy $S_{\text{LE}}=-\sum_n p_n \ln p_n$,
where $p_n=|\langle \phi_n |\psi_0\rangle|^2$ are provided. These data were calculated using the TLFFA algorithm with $L=80$ and 8192 truncated states for several values of $\kappa$.  
It can be observed that as $\kappa$ decreases to $\kappa < 1$, the number of peaks increases, accompanied by spectral weight growth. This suggests an increase in the LE spectral entropy and, consequently, an enhancement of the many-body Landau-Zener tunneling due to the quantum geometric effects.  

\begin{figure}[t]
    \includegraphics[width=8cm]{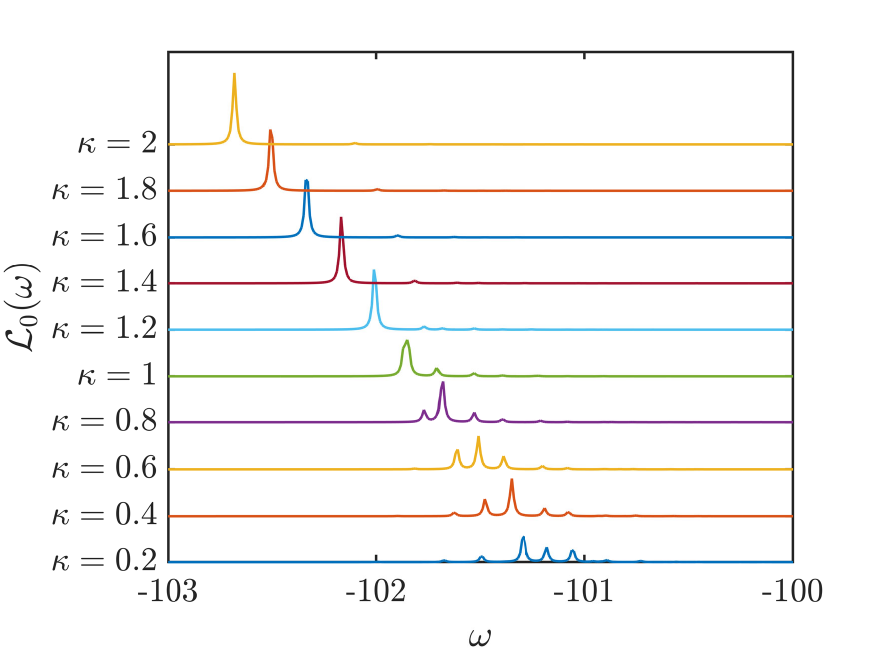}
    \caption{$-p \ln p$ $vs.$ $\w$ for different $\kappa$ from $0.2$ to $2$ with 
    $\Delta\kappa=0.2$, corresponding to the curves from the bottom to the top, after a Lorentzian broadening with $\alpha=0.001$.}
    \label{fig:suppNLEw}
\end{figure}

\end{document}